\documentclass[twocolumn,pre,aps]{revtex4}
\usepackage{graphicx}%
\usepackage{dcolumn}
\usepackage{longtable}%
\begin{document}


\title{Application of Minimal Subtraction Renormalization to Crossover
Behavior near the $^3$He Liquid-Vapor Critical Point}
\author{Fang Zhong, M. Barmatz, and Inseob Hahn}
\affiliation {Jet Propulsion Laboratory, California Institute of
Technology, Pasadena, California 91109-8099}
\date{\today}

\begin{abstract}
Parametric expressions are used to calculate the isothermal
susceptibility, specific heat, order parameter, and correlation length
along the critical isochore and coexistence curve from the asymptotic
region to crossover region.  These expressions are based on the
minimal-subtraction renormalization scheme within the $\phi^4$ model. 
Using two adjustable parameters in these expressions, we fit the theory
globally to recently obtained experimental measurements of isothermal
susceptibility and specific heat along the critical isochore and
coexistence curve, and early measurements of coexistence curve and
light scattering intensity along the critical isochore of $^3$He near
its liquid-vapor critical point.  The theory provides good agreement
with these experimental measurements within the reduced temperature
range $|t| \le 2\times 10^{-2}$.
\end{abstract}

\maketitle

\section{Introduction}
It is well known that thermodynamic quantities exhibit singularities
asymptotically close to the critical point.  The power-law behavior of
these singularities, characterized by critical exponents and the
concept of universality and scaling, have been successfully described
by renormalization-group (RG) theory.  Away from the asymptotic region,
thermodynamic quantities of real physical systems deviate from simple
power-law behavior. However, RG theory can still provide insight in
understanding critical crossover behavior.  

There are two main field-theoretical renormalization-group schemes to
treat critical-to-classical crossover phenomena.  Dohm and co-workers
developed the minimal-subtraction renormalization (MSR) scheme
\cite{schloms89} while Bagnuls and Bervillier developed the massive
renormalization (MR) scheme \cite{bagnuls85}.  Both of these theories
used the Borel resummation technique to describe the crossover behavior
of the $\phi^4$ model in any $O(n)$ universality class and in three
dimensions.  The difference between the two schemes was discussed in
ref.\cite{schloms89}. These field-theoretical crossover theories were
improved over the years as asymptotic theories became more accurate
\cite{guida98}.  Recently, Larin {\it et al}. improved the MSR
expressions for the specific heat and compared their results with the
superfluid helium $(n=2)$ system \cite{larin98}.  Bagnuls and Bervillier
have also improved their theory to match the recent asymptotic values
for exponents and leading amplitude ratios \cite{bagnuls02}.  Both
renormalization schemes can provide crossover functional forms for
thermal properties with a minimal set of fluid-dependent adjustable
parameters.  However, a direct comparison between these recent
theoretical predictions and different experimental measurements near the
liquid-vapor critical point $(n=1)$ has been lacking.

In this paper we will present a direct comparison between the MSR field
theoretical crossover functions and various experimental measurements
near the liquid-gas critical point of $^3$He.   The comparison using
the MR theory will be published elsewhere.   

The paper is divided into two parts.  In the first part, we briefly
summarize the MSR functional expressions for susceptibility, specific
heat, coexistence curve, and correlation length from previous work
\cite{schloms89,schloms90,krause90,halfkann92}.  In addition, we
derived within the MSR framework new RG functional expressions for the
asymptotic critical amplitudes of the susceptibility and coexistence
curve as well as the first coefficient in a Wegner expansion for
susceptibility, specific heat, coexistence curve, and correlation
length.  From these expressions, universal amplitude ratios for the
$O(1)$, 3-dimensional system are calculated and compared with the most
recent values from Bagnuls {\it et al}. \cite{bagnuls02} and Fisher
{\it et al}. \cite{fisher98}.

The second part of the paper includes the results of MSR functional
fits to experimental measurements.  In our previous work, we analyzed
the isothermal susceptibility of $^3$He along the critical isochore
above $T_c$ using theoretical expressions based upon the
minimal-subtraction scheme \cite{hahn01}.  In this work, we combine that
analysis with susceptibility measurements along the coexistence curve
and specific heat measurements along the critical isochore
\cite{barmatz00}. Measurements of coexistence curve and the light
scattering intensity  near the critical point of $^3$He
\cite{luijten02, miura84} are also analyzed.

\section{Theoretical Expressions}
The Hamiltonian for the $\phi^4$ model in three dimensions $(d=3)$ is
\begin{equation}\label{eq:Hamiltonian}
H_\phi = \int d^3 x \bigl\{ {\textstyle{1\over 2}} r_0 \phi_0^2 +
{\textstyle{1\over 2}} (\nabla\phi_0)^2 + u_0 \phi_0^4 \bigr\} \,,
\end{equation}
where $\phi_0$ is the order parameter field, whose statistical mean
value is the physical order parameter of a given system.   The parameter
$u_0$ is the fourth-order coupling constant.  The parameter $r_0$ is
related to the reduced temperature $t\equiv(T-T_c)/T_c$ by
\begin{equation}\label{eq:coupling_r0}
r_0 = a_0 \, t\,,
\end{equation}
where $a_0$ is a nonuniversal constant.  It is important to note that
the total Hamiltonian is the sum of $H = H_\phi + H_0$, where $H_0$ is
the analytic background free energy.  Since the liquid-vapor critical
point has a single component order parameter and belongs to the 
$O(1)$ universality class, we have $n=1$. 

The dimensionless bare order parameter field $\phi_0$ and the bare
coupling parameters $u_0$ and $r_0$ are renormalized to
[Eqs.~(S2.11)and (S2.12)
\cite{schloms90}]
\begin{eqnarray}
\phi &=& Z_\phi(u,\epsilon)^{-1/2}\,\phi_0
\label{eq:renormalize_phi} \\ u &=& \mu^{-1}\,
Z_u(u,\epsilon)^{-1}\, Z_\phi(u,\epsilon)^2\, A_3\, u_0
\label{eq:renormalize_u} \\ r &=& a \,
t = Z_r(u,\epsilon)^{-1} \, a_0 \, t \label{eq:renormalize_r} \,,
\end{eqnarray}
where $A_3 = (4\pi)^{-1}$ is a geometric factor and $\epsilon = 4-d =
1$ for dimension $d=3$.  The $Z$ factors are associated with their
respective field-theoretic functions \cite{schloms89}
\begin{eqnarray}
\zeta_r(u) &=& \left. \mu \partial_\mu \ln Z_r(u,\epsilon)^{-1}
\right|_0 \,, \label{eq:zeta_r_0}\\
\zeta_\phi(u) &=& \left. \mu \partial_\mu \ln Z_\phi(u,\epsilon)^{-1}
\right|_0 \,, \label{eq:zeta_phi_0}\\
\beta_u(u) &=& u\left[-1 + \left. \mu \partial_\mu
\left(Z_u^{-1}Z_\phi^2\right) \right|_0 \right] \,, \label{eq:beta_u_0}
\end{eqnarray}
where the index 0 means differentiation at fixed $r_0$, $\phi_0$,
and $u_0$.  

By introducing a flow parameter $l$, the effective coupling
${u}(l)$ satisfies the flow equation
\begin{equation}\label{eq:flow_eqn_u}
l{{d{u}(l)}\over{dl}} = \beta_u({u}(l))\,.
\end{equation}
The flow parameter $l$ is related to the correlation length by
\begin{equation}\label{eq:l2xi}
\xi(l) = (\mu\,l)^{-1}\,,
\end{equation}
with $\mu^{-1}$ being an arbitrary reference length.  The flow
parameter $l = 0$ corresponds to the Ising fixed point ${u}(l=0) = u^*$,
which is determined from $\beta_u(u^*) = 0$.  The effective coupling
${r}(l)$ satisfies the flow equation
\begin{equation}\label{eq:flow_eqn_r}
l{{d{r}(l)}\over{dl}} = {r}(l)\zeta_r({u}(l))\,.
\end{equation}
The flow parameter $l = 1$ is an arbitrary reference point, at which
the nonuniversal initial values are ${u}(l=1) = u$ and ${r}(1) = r =
a\,t$. 

The field-theoretic functions $\zeta_r(u),\zeta_\phi(u)$ and
$\beta_u(u)$ in Eqs.~(\ref{eq:zeta_r_0})-(\ref{eq:beta_u_0}) are known
up to five-loop order in perturbation expansions around $u = 0$
\cite{schloms89}.  However, the expansions do not converge away from
$u=0$.  Hence Borel resummations were used on the expansions to
calculate the values of these functions over the range $0 < u \le u^*$. 
For most investigations, see for instance
ref.\cite{schloms90,krause90,halfkann92}, only the function values at
the fixed point $u^*$ were calculated using the Borel resummations on
the five-loop expansions.  The function values over the range $0 < u < 
u^*$ were obtained using up to two-loop order expressions with
extrapolation terms added in order to reproduce the values at the fixed
point \cite{schloms89}.  For a system of dimension $d=3$ and single
component order-parameter $n=1$, one obtains
\begin{eqnarray}
\zeta_r(u) &=& 12u - 120u^2 + a_1 u^3 - a_2 u^4\,, \label{eq:zeta_r}\\
\zeta_\phi(u) &=& -24 u^2 + a_3 u^3 \,, \label{eq:zeta_phi}\\
\beta_u(u) &=& - u + 36 u^2 (1+a_4u)/(1+a_5u)\,. \label{eq:beta_u}
\end{eqnarray}
Here $a_1$ through $a_5$ are the coefficients for the extrapolation
terms with values listed in Appendix \ref{dix:const}.  Using these
functions and the flow equations, thermal properties along the
critical isochore and coexistence curve can be calculated from the
asymptotic to crossover regions using the initial values for
Eqs.~(\ref{eq:flow_eqn_u}) and (\ref{eq:flow_eqn_r}), $u = u(l=1)$ and
$a = a(l=1)$, and the arbitrary length scale $\mu^{-1}$ in
Eq.~(\ref{eq:l2xi}).

\subsection{Reduced temperature}

Within the MSR scheme, the expression for the reduced temperature in
terms of the flow parameter $l$ can be derived as follows.  The reduced
temperature $t$ and the flow parameter $l$ can be linked using
Eqs.~(S4.25), and (S4.26) in ref.\cite{schloms89} and Eq.~(H2.9) of
ref.\cite{halfkann92}, together with the solution of
Eq.~(\ref{eq:flow_eqn_r})
\begin{eqnarray}\label{eq:t2la}
\nonumber
{r}(l) &=& {r}(1) \exp \int\limits_1^l {\zeta
_r{{dl^\prime} \over {l^\prime}}} \\ \nonumber
 &=& a\left| t \right|\exp \int\limits_1^l {\zeta
_r{{dl^\prime} \over {l^\prime}}} \\ &=& b_\pm ({u}(l))\, \mu
^2l^2\,,
\end{eqnarray}
with
\begin{eqnarray}
b_+({u}(l)) &=& Q({u}(l))\,,
\label{eq:bplus}\\
b_-({u}(l)) &=&
\textstyle{3\over 2} -  Q({u}(l)) \label{eq:bminus}\,.
\end{eqnarray}
Here `+' is for $T>T_c$ and `$-$' is for $T<T_c$.  Krause {\em et al}.
\cite{krause90} determined a one-loop expression plus a higher-order
extrapolation for $Q(u)$ given by [Eq.~(K3.5)]
\begin{equation}
Q(u) = 1 + b_Q\, u^2\, \ln(c_Q u)\,, \label{eq:Qu1loop}
\end{equation}
where $b_Q$ and $c_Q$ are the extrapolation coefficients with the
values given in Table~\ref{table:extrapolation} in
Appendix~\ref{dix:const}.

By adding and subtracting $\zeta_r^* = \zeta_r(u^*)$ in the integrand of
Eq.~(\ref{eq:t2la}) and using the identity $\nu^{-1} = 2 -\zeta_r^*$,
where $\nu$ is the critical exponent of correlation length $\xi$, one
arrives at
\begin{equation}\label{eq:t2lb}
a\left| t \right|\exp \int\limits_1^l {\left( \zeta _r - \zeta
_r^* \right) {{dl^\prime} \over {l^\prime}}}=b_\pm (l)\mu
^2l^{1/\nu}\,.
\end{equation}
By rearrange Eq.~(\ref{eq:t2lb}), one obtains the following expression
for the reduced temperature
\begin{equation}\label{eq:t2lc}
|t| = b_\pm(l) \, t_0 \, l^{1/\nu} \exp[-F_r(l)]\,,
\end{equation}
with
\begin{equation}\label{eq:t0}
t_0 = {\mu^2\over a} \exp[F_r(1)]\,,
\end{equation}
and
\begin{eqnarray}\label{eq:Fr}
F_r(l) &=& \int_0^l {d l^\prime \over l^\prime} [
\zeta_r({u}(l^\prime)) - \zeta_r(u^*)] \\ \nonumber
 &=& \int_{u^*}^{{u}(l)} { {\zeta_r(u^\prime) -
\zeta_r(u^*) }\over {\beta_u(u^\prime)}} d u^\prime  \,.
\end{eqnarray}

\subsection{Susceptibility}

\subsubsection{General expression}

The following expressions for the dimensionless susceptibility $\chi_T^*
\equiv
\chi_\pm$ were given in refs.~\cite{krause90,halfkann92} respectively
for $T > T_c$ [Eq.~(K2.7)] and $T < T_c$ [Eq.~(H2.16)],
\begin{equation}\label{eq:chi1}
\chi _\pm ={{Z_\phi (u)} \over {\mu ^2l^2f_\pm ({u}(l))}}\exp
\int\limits_l^1 {\zeta _\phi {{dl'} \over {l'}}} \,.
\end{equation}
The amplitude functions, $f_\pm$, were expressed to two-loop order plus
a higher-order extrapolation, [Eqs.~(K3.1) and (H4.2)], to give
{\setlength{\arraycolsep}{0pt}
\begin{eqnarray}\label{eq:amp_chit}
f_+[u] &=&  1 - {92\over 9} u^2 ( 1 + b_\chi u)
\,\,\,\,\,\,\,\,\,\,\,\,\,\, (T>T_c), \\  \nonumber f_-[u] &=& \left[1 -
18u + 159.56 u^2 ( 1 + d_\chi u)\right]^{-1} \, (T<T_c),\,
\end{eqnarray}}
where $b_\chi$ and $d_\chi$ are the extrapolation coefficients with the
values given in Table~\ref{table:extrapolation} in
Appendix~\ref{dix:const}.

The minimal renormalization factor, $Z_\phi$, in Eq.~(\ref{eq:chi1}) is
given by [Eq.~(K\,\,A12)]
\begin{equation}\label{eq:zphi}
Z_\phi(u)^{-1} = \exp \int\limits_{0}^u {d u^\prime {{\zeta
_\phi(u^\prime)} \over {\beta_u(u^\prime)}}}\,.
\end{equation}
The expression
\begin{equation}\label{eq:chi2}
\chi _\pm = \chi_0 \, l^{-\gamma/\nu}\, {{\exp
[-F_\phi({u}(l))]} \over {f_\pm ({u}(l))}}\,
\end{equation}
can be obtained by adding and subtracting $\zeta_\phi^* =
\zeta_\phi(u^*)$ in the integrand of Eq.~(\ref{eq:chi1}), and using the
relations $\zeta_\phi^* = -\eta$ \cite{schloms89} and $\gamma = \nu(2 -
\eta)$, where $\eta$ is the critical exponent of the fluctuation
correlation at the critical point and $\gamma$ is the critical exponent
of susceptibility.  In Eq.~(\ref{eq:chi2})
\begin{equation}\label{eq:chi_0}
\chi _0 = \mu^{-2}\, Z_\phi(u)\, \exp [F_\phi(1)] \,,
\end{equation}
and
\begin{eqnarray}\label{eq:Fphi}
F_\phi(l) &=& \int_0^l {d l^\prime \over l^\prime}
[ \zeta_\phi({u}(l^\prime)) - \zeta_\phi(u^*)] \\ \nonumber
 &=& \int_{u^*}^{{u}(l)} { {\zeta_\phi(u^\prime) -
\zeta_\phi(u^*) }\over {\beta_u(u^\prime)}} d u^\prime \,.
\end{eqnarray}
Using Eq.~(\ref{eq:l2xi}) and $\gamma = \nu(2 - \eta)$,
Eq.~(\ref{eq:chi2}) can be rewritten as
\begin{equation}\label{eq:chi2a}
\chi _\pm = \xi^{2-\eta}\, \chi_0 \, \mu^{2-\eta}\,{{\exp
[-F_\phi({u}(l))]} \over {f_\pm ({u}(l))}}\,.
\end{equation}
Thus, in the asymptotic regime ($F_\phi \rightarrow 0$)
\begin{equation}\label{eq:chi2b}
\chi _\pm = D\,\xi^{2-\eta}
\end{equation}
with a nonuniversal proportionality constant $D \equiv \chi _0\,
\mu^{2-\eta}\, f_\pm^{-1} (u^*)$.

\subsubsection{Critical Amplitudes}
Within the pure $\phi^4$ model, the standard Wegner expansion for the
susceptibility is given by
\begin{equation}\label{eq:chi_Wegner}
\chi _\pm = \Gamma_0^\pm |t|^{-\gamma}(1 + \Gamma_1^\pm\,
|t|^{\Delta} + \Gamma_2^\pm\, |t|^{2\Delta} + \cdot\cdot\cdot)\,,
\end{equation}
where $\Gamma_0^\pm$ are the leading asymptotic critical amplitudes, 
$\Gamma_1^\pm$ are the first Wegner expansion amplitudes above and
below the transition, and $\Delta$ is the correction-to-scaling
exponent.  The details of the derivations of the leading and first
Wegner critical amplitudes are given in Appendix
\ref{dix:chi_amplitudes}.  Here we list the derived expressions for the
critical amplitudes,
{\setlength{\arraycolsep}{1pt}
\begin{eqnarray}\label{eq:Gamma_0}
\Gamma_0^\pm &=& {{\chi_0 \, (b_\pm^* t_0)^\gamma}\over
{f_\pm(u^*)}} \,,
\\
\label{eq:Gamma_1} \Gamma_1^\pm &=&
\left.\left(\gamma{{\zeta_r^\prime}\over{\omega}} - \gamma
{{b_\pm^\prime}\over{b_\pm}} + {{\zeta_\phi^\prime}\over{\omega}}
+ {{f_\pm^\prime}\over{f_\pm}} \right)\right|_{u^*} {{u^* -
u}\over {\left(b_\pm^*\, t_0\right)^{\Delta}}}\,\,\,\,\,\,\,\,\,\,\,
\end{eqnarray}}
with $\Delta = \nu\omega$ and $\omega = d\beta_u/du|_{u^*}$.

\subsection{Specific Heat}
\subsubsection{General expressions}

The total specific heat is usually separated as
\begin{equation}\label{eq:c_total}
C^\pm = C_B + C_{\phi}^\pm\,,
\end{equation}
where the term $C_B > 0$ represents an analytic ``background"
contribution from the analytic background free energy $H_0$, and
$C_{\phi}^\pm$ represents the critical contribution from order
parameter fluctuations. Here `+' is for the specific heat above
$T_c$ along the critical isochore, `$-$' is for below $T_c$ in
coexisting phases.

The critical specific heat $C_{\phi}^\pm$ derived from the
Hamiltonian expressed in Eq.~(\ref{eq:Hamiltonian}) has two
representations within the MSR scheme.  These two representations
are derived via multiplicative and additive renormalization as
detailed in ref.~\cite{schloms90}.  The most recent work by Larin
{\it et al}.\cite{larin98} used the representation via additive
renormalization that we will use in this paper.

The critical specific heat $C_{\phi}^\pm$ per unit volume near $T_c$ is
expressed by [Eqs.~(S2.36) or (L3.3)]\cite{schloms90,larin98}
{\setlength{\arraycolsep}{1pt}
\begin{eqnarray}\label{eq:C_bare}
C_{\phi}^\pm &=&T_c^2V^{-1}{{\partial ^2} \over {\partial T^2}}\ln
\int {D\phi\exp -H_\phi } \\ \nonumber
  &=&{\textstyle{1\over 4}} a^2
\mu^{-1} A_3 K_\pm({u}(l))  \exp\int_u^{{u}(l)} {{{
{2\zeta _r({u}^\prime) -1}} \over {\beta_u({u}^\prime)}}
d{u}^\prime} \,.
\end{eqnarray}}
The amplitude functions $K_\pm(u)$ are given by
\begin{equation}\label{eq:Ku}
K_\pm(u) = F_\pm(u) - A(u)\,.
\end{equation}
The functions $F_\pm\left(u\right)$ for $n=1$ can be expressed by a
two-loop calculation plus a higher-order extrapolation,
[Eqs.~(K3.4) \cite{krause90} and (H4.4)\cite{halfkann92}]
\begin{equation}\label{eq:amp_cv}
F_\pm[u] = \cases{ -1 -6u(1+b_F u) & ($T>T_c$) \cr (2u)^{-1}
-4(1+d_F u) & ($T<T_c$) \cr }\,,
\end{equation}
where $b_F$ and $d_F$ are the extrapolation coefficients with the
values given in Table~\ref{table:extrapolation} in
Appendix~\ref{dix:const}.  The function $A(u)$ in Eq.~(\ref{eq:Ku}) is
governed by
\begin{equation}\label{eq:Au_diff}
l{{dA({u}(l))} \over {dl}}= 4B({u}(l)) + \left\{1 - 2\zeta
_r( {u}(l)) \right\} A({u}(l))\,,
\end{equation}
with $A(u=0) = -4B(u=0)$.  The function $B(u)$ has been calculated to
$O(u^5)$ for any given $n$ [Eq.~(L2.21)] \cite{larin98}.  However the
five-loop Borel resummation of $B(u)$ was only performed for $n = 1$ at
$u^*$.  Hence a new extrapolation term with a coefficient $b_B$ is added
to the two-loop expression \cite{larin98} in order to satisfy $B(u^*,
n=1)$,
\begin{equation}\label{eq:B_u}
B(u)={\textstyle{1\over 2}} + 9 (1+b_B u) u^2 \,.
\end{equation}
At the fixed point ${u}=u^*$, $ldA({u}(l))/dl = \beta_u(u) dA({u})/d{u}
= 0$ since $\beta_u(u^*) = 0$, and Eq.~(\ref{eq:Au_diff}) leads to
\begin{equation}\label{eq:A_star}
A^*\equiv A(u^*)= -\frac{4\nu B(u^*)}{\alpha}\,,
\end{equation}
where $2\zeta_r^* - 1 \equiv -\alpha/\nu$ \cite{schloms89} is used with
$\alpha$ being the critical exponent for specific heat at constant
volume.

The integral in the exponential of Eq.~(\ref{eq:C_bare}) can be
rewritten as
\begin{equation}\label{eq:zeta_r_integrate}
\int_1^l {{{dl'} \over {l'}}}\left( {2\zeta _r-1}
\right)=2\int_1^l {{{dl'} \over {l'}}}\left( {\zeta _r-\zeta _r^*}
\right)+\ln l^{-\alpha /\nu }
\end{equation}
using $2\zeta_r^* - 1 \equiv -\alpha/\nu$.  The expression
for the specific heat from the additive renormalization can now be
rewritten as
{\setlength{\arraycolsep}{1pt}
\begin{eqnarray}\label{eq:C_bare_2}
\nonumber C_{\phi}^\pm &=& {{a^2} \over {16\pi \mu}} K_\pm(l)
l^{-\alpha /\nu }\exp \left[ 2F_r(l) \right] \exp \left[-2F_r(1)
\right]
\\  &=& C_0 l^{-\alpha /\nu } \exp \left[ 2F_r(l)
\right]K_\pm({u}(l)) \,,
\end{eqnarray}}
where $F_r(l)$ is given by Eq.~(\ref{eq:Fr}) and $C_0$ is
defined as
\begin{equation}\label{eq:C_0}
C_0 \equiv {{a^2} \over {16\pi \mu}}\exp \left[{-2F_r(1)}\right] =
\frac{\mu^3}{16\pi t_0^2}\,.
\end{equation}

\subsubsection{Critical Amplitudes}
The standard Wegner expansion within the pure $\phi^4$ model for
specific heat can be written as
\begin{eqnarray}\label{eq:cv_Wegner}
\nonumber {C^{\pm}} &=& A_0^\pm |t|^{-\alpha}(1 +
A_1^\pm|t|^\Delta + A_2^\pm|t|^{2\Delta} + \cdot\cdot\cdot)\\ &+&
B_{cr} +  C_B
\end{eqnarray}
where $B_{cr}$ is a constant background induced by long-range
correlations between the fluctuations.  The experimentally measured
constant background is the sum of $B_{cr}$ and the analytic
background, $C_B$.

The expression for the Wegner expansion of the specific heat via
multiplicative renormalization was derived and given in Eqs.~(S4.23)
and (S4.24) of ref.~\cite{schloms90}.  In
Appendix~\ref{dix:cv_amplitudes}, we derive the expressions for the
critical amplitudes and the critical background, $B_{cr}$, for the
representation via additive renormalization, using the technique that
is consistent with the one used for susceptibility. The results of
these derivations are,
{\setlength{\arraycolsep}{1pt}
\begin{equation}\label{eq:A_0}
A_0^\pm = {C_0 \, (b_\pm^* t_0)^\alpha} \left(F_\pm^* -A^* \right)
\,,
\end{equation}}
{\setlength{\arraycolsep}{1pt}
\begin{eqnarray}\label{eq:A_1}
\nonumber A_1^\pm &=& \left[ \frac{1}{A^* -F_\pm^*}
\left( F_\pm^\prime - \frac{2\nu}{\Delta - \alpha}(2B^\prime -
A^* \zeta_r^\prime) \right) \right. \,\,\,\,
\\   &-& \left.\left.
(2-\alpha)\frac{\zeta_r^\prime}{\omega}
  - \alpha \frac{b_\pm^\prime}{b_\pm} \right]\right|_{u^*} {{u^* -
u}\over {\left(b^*_\pm\, t_0\right)^{\Delta}}}\,,
\end{eqnarray}}
\begin{equation}\label{eq:Bcr_full}
-\frac{B_{cr}}{C_0} = A(u) - A^* + \frac{2\nu}{\Delta -
\alpha}(2B^\prime - A^* \zeta_r^\prime) (u^* - u)\,.
\end{equation}
The variables with a prime in Eqs.~(\ref{eq:A_1}) and
(\ref{eq:Bcr_full}) are derivatives with respect to $u$.  The right hand
side of Eq.~(\ref{eq:Bcr_full}) is negative for any given $u$.  Hence
one has $B_{cr} < 0$ since $C_0 > 0$ from Eq.~(\ref{eq:C_0}).

\subsection{Coexistence curve}
\subsubsection{General expressions}

In the liquid-vapor coexisting phases below $T_c$, the density
difference $\Delta \rho_{L,V} \equiv \rho_{L,V}/\rho_c - 1$ is the
statistical mean of the order parameter field $\langle \phi \rangle$. 
There is no asymmetry between $\Delta \rho_L$ and $\Delta \rho_V$
within the $\phi^4$ model. Schloms and Dohm have given an expression for
the square of the physical order parameter [Eq.~(S3.10)]
\cite{schloms90},
\begin{equation}\label{eq:phisqr}
\langle \phi \rangle^2 = A_3 Z_\phi(u) f_\phi(u(l_-))
\xi_-^{-1} \exp \int_{l_-}^1 \zeta_\phi
\frac{dl^\prime}{l^\prime}\,.
\end{equation}
Here $Z_\phi(u)$ is given in Eq.~(\ref{eq:zphi}). The amplitude function
$f_\phi(u)$ is expanded in one-loop with an extrapolation term,
[Eq.~(H4.1)]\cite{halfkann92}, to yield
\begin{equation}\label{eq:fphi}
f_\phi(u) = (8u)^{-1}(1 + d_\phi u)\,.
\end{equation}
The correlation length below $T_c$ is linked to the flow parameter
by $l_- = (\mu \xi_-)^{-1}$.  By combining these expressions
and following the derivation of Eq.~(\ref{eq:chi2}), one has
\begin{equation}\label{eq:phisqr2}
\langle \phi \rangle^2 = \phi_0^2\, l_-^{2\beta/\nu} f_\phi(
u(l_-)) \exp [-F_\phi(l_-)]\,,
\end{equation}
where $1+\eta = 2\beta/\nu$ is used and
\begin{equation}\label{eq:phi0sqr}
\phi_0^2 = (4\pi)^{-1}\mu Z_\phi(u) \exp [F_\phi(1)]\,,
\end{equation}
with $\beta$ being the critical exponent of the order parameter.

\subsubsection{Critical Amplitudes}

Using Eq.~(\ref{eq:t2lc}) to replace $l_-^{2\beta/\nu}$ in
Eq.~(\ref{eq:phisqr2}) and the scaling relations $\gamma =\nu(2-\eta)$,
$\alpha = 2-3\nu$, and $\alpha + 2\beta + \gamma = 2$, one has
{\setlength{\arraycolsep}{1pt}
\begin{eqnarray}\label{eq:phi}
\nonumber \langle \phi \rangle &=& \pm \phi_0 t_0^{-\beta}
|t|^\beta\,[b_-(l_-)]^{-\beta} [f_\phi(l_-)]^{1/2} \\ &\times&
\exp[\beta F_r(l_-)]\exp [-F_\phi(l_-)/2]\,.
\end{eqnarray}}
Expanding $b_-(l_-)$, $f_\phi(l_-)$, $F_r(l_-)$, and $F_\phi(l_-)$
in the same manner as described in
Appendix~\ref{dix:chi_amplitudes}, one obtains the Wegner expansion for
coexistence curve,
\begin{equation}\label{eq:coex_Wegner}
\Delta\rho_{L,V} \,  = \pm B_0 |t|^\beta (1 + B_1\, |t|^\Delta)
\,,
\end{equation}
with the leading critical amplitude and the first Wegner amplitude
being respectively
{\setlength{\arraycolsep}{1pt}
\begin{eqnarray}\label{eq:B_0}
B_0 &=& \phi_0 (b_-^*t_0)^{-\beta} \, (f_\phi^*)^{1/2}
\\\label{eq:B_1} B_1 &=&\left.
\left(\beta\frac{b_-^\prime}{b_-} - \beta\frac{\zeta_r^\prime}{\omega}
 - \frac{f_\phi^\prime}{2f_\phi} + \frac{\zeta_\phi^\prime}{2\omega}
 \right)\right|_{u^*} \frac{u^* - u}{\left(b_-^* t_0\right)^{\Delta}}
.\,\,\,\,\,\,\,\,\,\,\,\,
\end{eqnarray}}

\subsection{Correlation length}

Using Eq.~(\ref{eq:t2lc}) to express $l$ in terms of $|t|$, the
expression for dimensionless correlation length (if $\mu^{-1}$ is
taken dimensionless) is derived from Eq.~(\ref{eq:l2xi}) as
{\setlength{\arraycolsep}{1pt}
\begin{equation}\label{eq:correlation_length}
\xi \, |t|^{\nu} = \mu^{-1}\, \left[b_\pm({u}(l))
t_0\right]^{\nu} \, \exp[-\nu F_r(l)] \,.
\end{equation}}
An expansion of Eq.~(\ref{eq:correlation_length}) around $u(l)
\sim u^*$ to ${O[(u - u^*)^2]}$ leads to
{\setlength{\arraycolsep}{1pt}
\begin{eqnarray}\label{eq:xi_expanded}
\xi _\pm \, |t|^\nu &=& \mu^{-1} \left(b_\pm^* \, t_0\right)^\nu
\\ \nonumber &\times& \left.\left[1 + \nu\left(
{{\zeta_r^\prime}\over{\omega}} - {{b_\pm^\prime}\over{b_\pm}}
\right)\right|_{u^*} {{u^* - u}\over {(b_\pm^* t_0)^{\Delta}}}
|t|^{\Delta}\right]\,.
\end{eqnarray}}
This equation is identical to Eq.~(S4.8) of ref.\cite{schloms90}.
By comparing Eq.~(\ref{eq:xi_expanded}) to the standard Wegner
expansion form,
\begin{equation}\label{eq:xi_Wegner}
\xi \, |t|^\nu = \xi_0^\pm(1 + \xi_1^\pm\, |t|^\Delta) \,,
\end{equation}
one obtains the leading amplitudes and first Wegner correction
amplitudes of the correlation length
\begin{eqnarray}\label{eq:xi_0}
\xi_0^\pm &=& \mu^{-1} \left(b_\pm^* \, t_0\right)^\nu\,, \\
\label{eq:xi_1} \xi_1^\pm &=& \left.
\nu\left({{\zeta_r^\prime}\over{\omega}} -
{{b_\pm^\prime}\over{b_\pm}}\right)\right|_{u^*} {{u^* - u}\over
{\left(b_\pm^*\, t_0\right)^{\Delta}}}\,.
\end{eqnarray}

\subsection{Universal amplitude ratios}

Even though the leading amplitude and subsequent Wegner expansion
coefficients are fluid-dependent, certain combination ratios of these
amplitudes are universal. From the equations for the first Wegner
amplitudes of the specific heat, susceptibility, coexistence curve, and
correlation length, one notices that the system-dependent part, $(u -
u^*)/(b_\pm^* t_0)^\Delta$, is the same in every expression. 
Therefore the ratio of any of these first Wegner amplitudes is
universal based on the MSR $\phi^4$ model.  These universal ratios have
been given for the specific heat in ref.~\cite{larin98}, [
Eqs.~(\ref{eq:ratio_A_0}), (\ref{eq:ratio_A_1}), and
(\ref{eq:A_0_xi_0})]. In this paper we derive the other universal
ratios based on the MSR $\phi^4$ model. From Eqs.~(\ref{eq:Gamma_0})
and (\ref{eq:Gamma_1}), one has the universal amplitude ratios for
susceptibility,
\begin{equation}\label{eq:ratio_Gamma_0}
{{\Gamma_0^+}\over{\Gamma_0^-}} =
\left({{b_+^*}\over{b_-^*}}\right)^\gamma {{f_-(u^*)}\over{f_+(u^*)}}
= 4.94 \,,
\end{equation}
\begin{equation}\label{eq:ratio_Gamma_1}
{{\Gamma_1^+}\over{\Gamma_1^-}} =
\left({{b_-^*}\over{b_+^*}}\right)^\Delta
\left. {{\gamma{{\zeta_r^\prime}\over{\omega}} - \gamma
{{b_+^\prime}\over{b_+}} +
{{\zeta_\phi^\prime}\over{\omega}} + {{f_+^\prime}\over{f_+}}
}\over{\gamma {{\zeta_r^\prime}\over{\omega}} - \gamma
{{b_-^\prime}\over{b_-}} +
{{\zeta_\phi^\prime}\over{\omega}} + {{f_-^\prime}\over{f_-}}
}} \right|_{u^*} = 0.228 \,.
\end{equation}
From Eqs.~(\ref{eq:A_0}) and
(\ref{eq:A_1}), one obtains the universal amplitude ratios for
specific heat,
\begin{equation}\label{eq:ratio_A_0}
{{A_0^+}\over{A_0^-}} =
\left({{b_+^*}\over{b_-^*}}\right)^\alpha {{4\nu B^* +
\alpha F_+^*}\over{4\nu B^* +
\alpha F_-^*}} = 0.535 \,,
\end{equation}
\begin{equation}\label{eq:ratio_A_1}
{{A_1^+}\over{A_1^-}} =  1.07\,.
\end{equation}

Use of the scaling relation $\alpha + 2\beta + \gamma = 2$ and the
combination of Eqs.~(\ref{eq:Gamma_0}), (\ref{eq:A_0}), and
(\ref{eq:B_0}) leads to a universal ratio
{\setlength{\arraycolsep}{1pt}
\begin{eqnarray}\label{eq:ratio_Rc}
R_c &=& \frac{\alpha A_0^+ \Gamma_0^+}{B_0^2}
\\\nonumber &=& \frac{\alpha C_0 \chi_0 (F_+^* - A^*)(b_+^*t_0)^{\alpha +
\gamma}}{\phi_0^2(b_-^*t_0)^{-2\beta}f_\phi^*f_+^*}\\ \nonumber
&=& \frac{\alpha (b_+^*)^{\alpha +\gamma} (b_-^*)^{-2\beta}(F_+^*
- A^*)}{4f_\phi^*f_+^*} = 0.0580\,.
\end{eqnarray}}

From Eqs.~(\ref{eq:xi_0}) and (\ref{eq:xi_1}), the universal amplitude
ratios for the correlation length are,
\begin{equation}\label{eq:ratio_xi_0}
{{\xi_0^+}\over{\xi_0^-}} =
\left({{b_+^*}\over{b_-^*}}\right)^\nu  = 1.42 \,,
\end{equation}
\begin{equation}\label{eq:ratio_xi_1}
{{\xi_1^+}\over{\xi_1^-}} =  \left({{b_-^*}\over{b_+^*}}\right)^\Delta
\left. \frac{\frac{\zeta_r^\prime}{\omega} -
\frac{b_+^\prime}{b_+}} {\frac{\zeta_r^\prime}{\omega} -
\frac{b_-^\prime}{b_-}} \right|_{u^*} = 1.10\,.
\end{equation}

From Eqs.~(\ref{eq:A_0}) and (\ref{eq:xi_0}), one has the
universal relation between the amplitude of specific heat and
correlation length
{\setlength{\arraycolsep}{1pt}
\begin{eqnarray}\label{eq:A_0_xi_0}
\nonumber
\alpha\,A_0^\pm\,(\xi_0^\pm)^3 &=& \mu^{-3} C_0
(b_\pm^* t_0)^{\alpha + 3\nu} (4\nu B^* + \alpha F_\pm^*) \\ \nonumber
&=& \frac{1}{16\pi} \left(b_\pm^*\right)^2 (4\nu B^* +
\alpha F_\pm^*) \\  &=& \cases{ 0.0206 & ($T>T_c$) \cr 0.0134 &
($T<T_c$)
\cr }\,,
\end{eqnarray}}
where the scaling relation $\alpha + 3\nu = 2$ has been used. 
Equation~(\ref{eq:A_0_xi_0}) is identical to Eq.~(S4.22) of
ref.\cite{schloms90}.  The evaluation of the right-hand side uses the
constants given in Appendix~\ref{dix:const}.  A natural extension of
Eq.~(\ref{eq:A_0_xi_0}) is the universal relation between specific heat
and the correlation length throughout the crossover region.  Using
Eqs.~(\ref{eq:l2xi}), (\ref{eq:t2lc}), and (\ref{eq:C_bare_2}), and the
scaling relation $\alpha + 3\nu = 2$, one has
\begin{equation}\label{eq:cvxiuniversal}
C_\phi^{\pm} \xi_\pm^3 = \frac{b_\pm^2(l)\,K_\pm(l)}{16\pi |t|^2}\,.
\end{equation}
Since there is no fluid-dependent parameters appearing on the right
hand side of Eq.~(\ref{eq:cvxiuniversal}), the product of the critical
specific heat and the cubic of the correlation length is universal for
any given temperature throughout the crossover region.

Table \ref{table:ratios} lists the various universal amplitude ratios
derived from the minimal-subtraction renormalization scheme, Bagnuls
and Bervillier's massive-renormalization scheme \cite{bagnuls02}, and
other methods, such as $\varepsilon$-expansion, summarized by Fisher
and Zinn \cite{fisher98}.  The values given by Bagnuls and Bervillier
are closely matched to the values given by Guida and Zinn-Justin
\cite{guida98} after the readjustment of the Borel resummation criteria
\cite{bagnuls02}.  Noticeable differences exist in Table
\ref{table:ratios} among various theories.   In attempting to explain
these differences, two factors are important to note.  First, we are
unable to evaluate the uncertainties of the universal ratios since the
uncertainties on the Borel resummations at the fixed point $u^*$ for
most of the amplitude functions were not given in previous studies. 
Secondly, Eqs.~(\ref{eq:Gamma_1}), (\ref{eq:A_1}), and (\ref{eq:xi_1})
use the derivatives of Eqs.~(\ref{eq:zeta_r}), (\ref{eq:zeta_phi}),
(\ref{eq:Qu1loop}), (\ref{eq:amp_chit}), (\ref{eq:amp_cv}), and
(\ref{eq:B_u}) which could have sizable systematic uncertainties. 
These equations were only obtained from two-loop calculations and
extrapolated to the five-loop fixed point values with adjustable
constants.  Hence it is desirable to have these derivatives calculated
at the fixed point with Borel resummations.  Then the extrapolation
coefficients can be more accurately reconstructed, leading to the
estimates of the first Wegner coefficients with less uncertainties.

\begin{table}
\caption{\label{table:ratios}The values of various universal amplitude
ratios.  The calculation for this work uses the values of the
amplitude functions at the fixed point $u^*$ given in
Table~\protect{\ref{table:fixedpoint}} and the values of the critical
exponents given by Guida and Zinn-Justin \cite{guida98}.}
\begin{ruledtabular}
\begin{tabular}{c|c|c|c} 
Amplitude \\ ratios & This work & B and B\footnote{Bagnuls and
Bervillier
\cite{bagnuls02}} & F and Z\footnote{Fisher and Zinn \cite{fisher98}}
\\ \hline
$\Gamma_0^+ / \Gamma_0^-$ & $4.94$ &
$4.79\pm 0.10$ & $4.95 \pm 0.15$\\
$A_0^+/A_0^-$ & $0.535$ & $0.537\pm 0.019$ & $0.523\pm 0.009$\\
$\xi_0^+ / \xi_0^-$ & $1.42$ & $ $ & $1.89\pm 0.01_5$\\\hline
$R_c$ & $0.0580$ & $0.0574\pm 0.0020 $ & $0.0581\pm 0.0010$\\
$\alpha\,A_0^+\,(\xi_0^+)^3$
& $0.0206$ & $0.0196\pm 0.0001$ & $0.0188\pm 0.0001_5$\\
$\alpha\,A_0^-\,(\xi_0^-)^3$ & $0.0134$ & $ $ & $0.0053\pm
0.0002_5$\\\hline
$\Gamma_1^+ / \Gamma_1^-$ & $0.228$ & $0.215\pm0.029$ & $ $\\
$A_1^+/A_1^-$ & $1.07$ & $1.36\pm 0.47$ & $ $\\
$\xi_1^+ / \xi_1^-$ & $1.10$ & $ $ & $ $\\
$B_1/\Gamma_1^+$ & $0.76$ & $0.40\pm 0.35$ & $ $\\
\end{tabular}
\end{ruledtabular}
\end{table}

\section{Fit to experimental measurements}

The expressions within the MSR model are parametric for
susceptibility, specific heat, coexistence curve, and correlation
length along the critical isochore and coexistence curve.  We made a
variable change of $l = \exp(-x)$ in solving those expressions
numerically.  $x$ was discretized with 1000 data points over the range
of $-\infty < x < \infty$ to obtain the solution for ${u}(x)$ over the
range of $0 \leq {u}(x) \leq u^*$.   For each thermal property versus
reduced temperature, say $\chi_+$ vs $t$,  1000 data points were
calculated for a look-up table of $\chi_+(x_i)$ vs $t(x_i)$ with $i=1,
\cdot\cdot\cdot 1000$. The intended property was then obtained for a
given reduced temperature $t$ using a cubic spline.

This MSR model has three system dependent parameters, $u$, $\mu$, and
$a$ which fix the scales for $u(l)$, $\xi(l)$, and $t(l)$ in
Eqs.~\ref{eq:flow_eqn_u}, \ref{eq:l2xi}, and \ref{eq:t2lb}.  Here
$u(l)$, $\xi(l)$, and $t(l)$ are defined implicitly as functions of the
RG flow parameter $l$.  Since $l$ is eliminated in final solution, it is
clear that one of the three amplitudes is redundant.  This should
not be mistaken as a minimal number of three fitting parameters for a
complete equation-of-state while we only fit the thermal properties
along the critical isochore and coexistence curve.  In this paper
$\{\mu, a\}$ are chosen as fitting parameters for a prefixed $u$
because their combination only appears in the amplitude of the
parametric expressions, such as $t_0$, $\chi_0$, $C_0$, etc.   The
$u$ value is chosen based on the consideration that the expressions for
the first Wegner amplitudes were derived by ignoring higher order terms
in $[{u}(l) - u^*]$.  Therefore an accurate determination of the first
Wegner amplitudes requires $u$ to be close to $u^*$.

Besides $\{\mu, a\}$, the critical temperature can also be
a fitting parameter.  Another adjustable parameter is required for the
analytic background contribution to specific heat.  In fitting the
experimental data, $y_{\rm expt}$, to theory, we minimize
\begin{equation}\label{eq:chisqr}
\chi ^2=\sum\limits_{i=1}^N {\left( {{{y_{\rm expt}(x_i)-y_{\rm
theory}(x_i,\vec a)} \over {\sigma _i}}} \right)}^2 \,.
\end{equation}
Here $\vec a$ is an array of fitting parameters with the standard
error $\sigma$ given by
\begin{equation}\label{eq:sigma}
\sigma ^2=\sigma _y^2+\left( {{{\partial y} \over {\partial x}}}
\right)_{\vec a}^2\sigma _x^2.
\end{equation}
The partial derivative in Eq.~(\ref{eq:sigma}) is evaluated
numerically in each fitting iteration. The $x$ in
Eq.~(\ref{eq:chisqr}) is temperature.  In our experiment, the sample
temperature was determined from a resistance measurement having an
approximately 10~$\mu$K uncertainty, i.e. $\sigma_x = 1\times
10^{-5}$~K.  In fitting the measurements of isothermal susceptibility
and specific heat, we assign $\sigma_y = k \times y / 100$, assuming a
$k$\% uncertainty in the measurement.

The goodness of a fit is characterized by the value
\begin{equation}\label{eq:fitgoodness}
\chi_\nu^2=\frac{\chi^2}{N-M}\,,
\end{equation}
where $N$ is the number of data points and $M$ is the number of
fitting parameters.

All the experimentally measured quantities were made dimensionless by
expressing them in units of appropriate combinations of the $^3$He
critical temperature $T_c = 3.315$~K, critical density $\rho_c =
0.04145 {\rm g/cm^3}$, and critical pressure $P_c = 1.14$~Pascal.

The experimental susceptibility, $\chi_T = \rho (\partial \rho/\partial
P)_T$, is scaled by $\rho_c^2/P_c$ to obtain the dimensionless
susceptibility, $\chi_T^*\equiv \chi_T P_c/\rho_c^2$.  The physical
order parameter, $\Delta\rho\equiv \rho/\rho_c -1$ is already
dimensionless.   The measured heat capacity had units of [C] = J/K.  It
was then divided by the fluid volume to have a unit of $[\rho C_V] =
{\rm J/(cm^3\,K)}$.  Since the energy unit is $\rm [J] = [P][V]$, a
dimensionless specific heat was obtained as $C_V^* \equiv  \rho C_V T_c
/ P_c$ with $P_c/T_c = 0.03463 {\rm J/(cm^3\,K)}$.

The critical specific heat $C_{\phi}^\pm$ per unit volume (divided
by Boltzmann's constant $k_B$) near $T_c$ is given by 
Eq.~(\ref{eq:C_bare_2}).  The volume scaling factor is $v_0 = k_B
T_c / P_c$, thus the length scaling factor is
\begin{equation}
l_{0} = v_0^{1/3} = (k_B T_c / P_c)^{1/3}\,.
\label{eq:l0}
\end{equation}
For $^3$He, one has $l_{0} = 7.36\, {\rm\AA}$.  It is assumed that
$\mu$ is dimensionless in the MSR $\phi^4$ model expressions with
$l_0^{-1}$ being the scaling factor, one defines $\xi^* \equiv
\xi/l_0$. 

\subsection{Fit to susceptibility measurements}

The susceptibility along the critical isochore $(\rho = \rho_c)$ was
determined using $PVT$ measurements from both sides around $\rho_c$.  
The susceptibility along the coexistence curve was also determined using
$PVT$ measurements.  For example, $\chi_T^{liq}$ was obtained from $PVT$
measurements for $\rho > \rho_{coex}^{liq}$ and $\chi_T^{vap}$ was
obtained from $\rho < \rho_{coex}^{vap}$. 

Since $\chi_T^*$ varies sharply as $\rho \rightarrow \rho_c$ and $\rho
\rightarrow \rho_{coex}$, the dominating uncertainty in $\chi_T^*$ comes
from the uncertainty in locating either $\rho_c$ for measurements above
$T_c$ and $\rho = \rho_{coex}$ for measurements below $T_c$.  Above
$T_c$, the inflection point was well confined by the data from the both
sides of $\rho_c$.  Below $T_c$, $\rho = \rho_{coex}$ was determined
from a kink in $P$ versus $\rho$ curve.   However, this kink becomes
less pronounced as $T \rightarrow T_c$.   Based on our observation, we
assigned the susceptibility uncertainties to be $\sigma_{\chi_T} (T
>T_c) = 0.02 {\chi_T} (T >T_c)$ and $\sigma_{\chi_T} (T <T_c) = 0.1
{\chi_T} (T <T_c)$.

\begin{figure}
\includegraphics[width=3.1in]{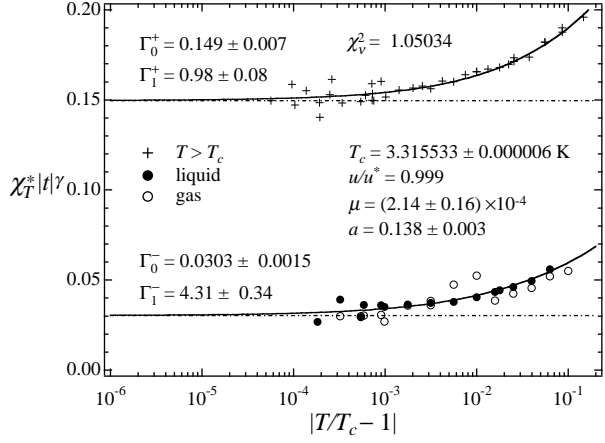}
\caption{Fit of MSR $\phi^4$ model to $^3$He susceptibility
measurements for both $T > T_c$ and $T < T_c$. $u/u^*$ is fixed to be
0.999, while $T_c$, $\mu$, and $a$ were adjusted. The solid line is the
best fit.   The dot-dashed straight lines represent the asymptotic
predictions from the fit.}
\label{fig:chi_T_leveled}
\end{figure}

The results of fitting the susceptibility measurements for both $T>T_c$
and $T < T_c$ to the MSR expression in Eq.~(\ref{eq:chi2}) is shown in
Fig.~\ref{fig:chi_T_leveled}.  The susceptibility was scaled by
$|t|^\gamma$ in order to provide a more sensitive representation of the
crossover behavior and the fitting quality.  The dot-dashed straight
lines represent the asymptotic predictions from the MSR fit.  The
uncertainties in the amplitudes were deduced from the uncertainties of
$\mu$ and $a$ in the fit.

\begin{figure}
\includegraphics[width=3.1in]{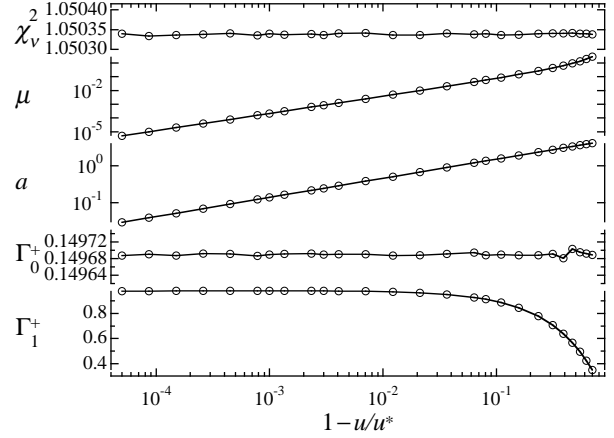}
\caption{The susceptibility fitting quality $\chi_\nu^2$, parameters
$\mu$ and $a$, and the resultant critical amplitudes versus fixed
$u/u^*$. $\Gamma_0^+$ and $\Gamma_1^+$ were calculated using
Eqs.~(\protect{\ref{eq:Gamma_0}}) and (\protect{\ref{eq:Gamma_1}}) for
each set of $\{u,\mu, a\}$ from the fit.}
\label{fig:Gamma1_vs_du}
\end{figure}

Figure~\ref{fig:Gamma1_vs_du} shows $\chi_\nu^2$ and $\{\mu, a\}$
versus $(1-u/u^*)$.  It is interesting to see that the goodness of the
fit remains unchanged over a wide range of $(1-u/u^*)$.  This verifies
that only two out of the three fluid-dependent parameters are relevant
fitting parameters.  We note here that no improvement in $\chi_\nu^2$
was made within scatters when $u/u^*$ was also free to be adjusted in
the fit.

Also shown in Fig.~\ref{fig:Gamma1_vs_du} are the calculated
$\Gamma_0^+$, and $\Gamma_1^+$ versus $(1-u/u^*)$.  $\Gamma_1^+$
reaches a plateau for $(1~-~u/u^*) \le 4\times 10^{-3}$ since the
analytical expression for $\Gamma_1^+$ was derived by ignoring
$O((1 - u/u^*)^2)$ or higher order terms.  Therefore in this work we
fix  the value of $u/u^* = 0.999$. 

Furthermore, $\mu \propto (1 - u/u^*)^{2\nu}$ and $a \propto (1 -
u/u^*)^{4\nu-2}$ for $(1-u/u^*) < 0.1$ implies a strong correlation
between $\mu$ and $a$ when susceptibility is chosen in fitting the MSR
$\phi^4$ theory to experimental data.  The above power-law dependence of
$\mu$ and $a$ on $(1-u/u^*)$ comes from the fact that $\Gamma_1^+$
is nearly a constant for $(1~-~u/u^*) < 1\times 10^{-2}$.  Thus from
Eq.~(\ref{eq:Gamma_1}) $t_0 \approx \mu^2/a \propto
(1~-~u/u^*)^{1/\Delta} \approx (1~-~u/u^*)^2$.

\subsection{Fit to specific heat measurements}

The specific heat, $C_V$, near the $^3$He critical point was measured
using a heat pulse method.  The temperature change could be measured
very accurately using a magnetic susceptibility thermometer with 1~nK
resolution.  For $T>T_c$, temperature equilibration was very fast due
to the ``piston effect'' \cite{behringer90}, and the uncertainty in the
measured $C_V$ was $\sim 1$\%, i.e. $\sigma_{C_V} (T >T_c) = 0.01 {C_V}
(T >T_c)$.  For $T < T_c$, equilibration underwent critical slowing down
as the fluid approached $T_c$.  The slowing-down was due to the mass
transfer at the meniscus between liquid and vapor.  Since the sample
cell was not perfectly adiabatic due to its mechanical support and
electrical wires, there was some heat loss from the cell to the
surrounding during the long equilibration.  The uncertainty in
measuring $C_V$ was typically 5\%, i.e. $\sigma_{C_V} (T <T_c) = 0.05
{C_V} (T <T_c)$.

\begin{figure}
\includegraphics[width=3.1in]{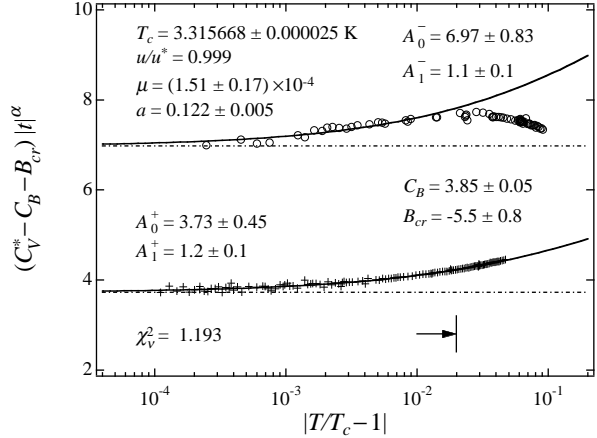}
\caption{The dimensionless specific heat at constant volume
versus reduced temperature.  The symbols represent the
experimental measurements.  The solid line is the best fit.  The
dot-dashed straight lines represent the asymptotic predictions
from the fit. The arrow indicates the fitting range $|t|\le
2\times 10^{-2}$.}
\label{fig:Cv_leveled}
\end{figure}

In fitting $C_V$ measurements to the MSR $\phi^4$ model, an additional
adjustable parameter, $C_B$, appears in Eq.~(\ref{eq:C_bare_2}).
By treating $C_B$ as a constant within a small reduced temperature
range around $T_c$, the true crossover behavior described by the
MSR $\phi^4$ model can be revealed.  Figure~\ref{fig:Cv_leveled}
shows a fit of the $C_V$ measurements for both $T>T_c$ and $T < T_c$ to
the MSR expression in Eq.~(\ref{eq:C_bare_2}).  The fit was limited to
the reduced temperature range $|t| \le 2\times 10^{-2}$ as indicated by
an arrow in the figure.   The agreement between the experimental
measurements and the theory is good.  The uncertainties in the critical
amplitudes and fluctuation-induced background were error-propagated
from the uncertainties of $\mu$ and $a$.

The fluctuation-induced background specific heat, $B_{cr}$, was
calculated from Eq.~(\ref{eq:Bcr}) in Appendix~\ref{dix:cv_amplitudes}.
Its absolute value is close to that of $C_B$. As a result, the combined
background specific heat is close to zero for $^3$He as first
demonstrated experimentally by Brown and Meyer
\cite{brown72a}.

\subsection{Fit to Coexistence Curve Measurements}

\begin{figure}
\includegraphics[width=3.10in]{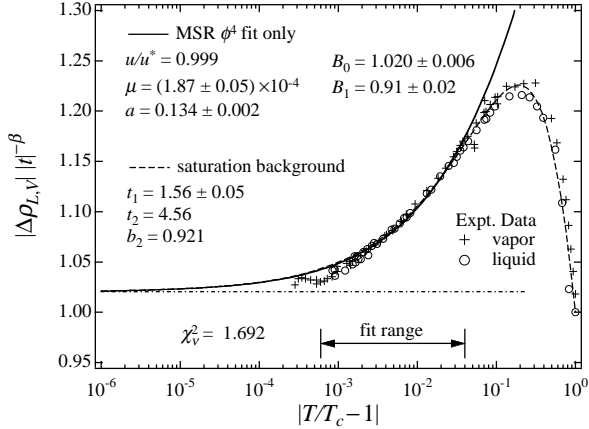}
\caption{Application of the MSR $\phi^4$ model to the data of the
$^3$He coexistence curve.  The solid line is the best fit with $\{\mu,
a\}$ adjusted.  The dashed line includes the empirical background
contribution with $\{\mu, a\}$ fixed from the fit without the
background.  The dot-dashed straight line represents the asymptotic
prediction from the fit.}
\label{fig:Coex_leveled}
\end{figure}

The best data for the $^3$He coexistence curve was compiled in a recent
paper by Luijten and Meyer \cite{luijten02}.  We apply the MSR $\phi^4$
model to the coexistence curve using these data as shown in
Fig.~\ref{fig:Coex_leveled}.   The fit was limited to the range
$6\times 10^{-4} \ge |t| \le 4\times10^{-2}$.  The lower bound was so
chosen since the measurements were affected by the gravity effect for
$|t| < 6\times10^{-4}$ due to a large cell height (4.3 mm) used in that
experiment \cite{pittman79}.  The upper bound was so chosen since the
$\phi^4$ model was developed for critical phenomena and did not include
analytic behavior associated with a system approaching absolute zero
temperature.  The standard deviation for $|\Delta\rho_{L,V}|$ was
approximated based on the percentile deviation in Fig.~5 of
ref.~\cite{pittman79}, namely 1\% at $|t| = 6\times10^{-4}$ and 0.2\%
at $|t| > 1\times10^{-2}$.  The standard deviation for reduced
temperature was $\delta T = 1\times 10^{-5}$K divided by $T_c =
3.3155$K.

The solid line in the figure represents the best fit with only $\{\mu,
a\}$ adjusted.   The predicted $B_0 = 1.02$ is consistent with the
reported $B_0 = 1.02$ in ref.~\cite{luijten02}.  The agreement between
the model calculation and the experimental data is satisfactory over
the fitting range.   The systematic difference between the MSR $\phi^4$
model calculation and the measurements over the fitting range also
exists from other theoretical model calculations \cite{luijten02}.

The systematic deviation between the MSR $\phi^4$ model calculation and
the experimental data over the fitting range may be due to the fact
that there was no proper background contribution included in the
analysis.  We attempt in this paper to include the effect of the order
parameter saturation as a possible background contribution. 
The saturation of order parameter at absolute zero temperature has been
studied by Povodyrev {\it et al}. \cite{anisimov99} for an ideal Ising
model.  We propose an empirical expression that is
consistent with that study for the limiting behavior at $|t|=1$.  Not
only does the order parameter saturate to a constant value but its
slope also approaches zero at $|t| = 1$.  In the case of the
liquid-vapor system where the physical order parameter is the
normalized density difference from the critical value, the saturation
value is also unity.  Our empirical expression, satisfying this
limiting behavior, is
\begin{equation}\label{eq:coex_bkgd}
\Delta \rho_{L,V} = \langle\phi\rangle \exp(-|t|/t_1) \pm b_2\left[1 -
\exp(-|t|/t_2)\right]\,.
\end{equation}
In fitting the expression in Eq.~(\ref{eq:coex_bkgd}) to the
experimental data, only $t_1$ is adjusted while $t_2$ and $b_2$ are
solved for a given $t_1$ through the constraints $\Delta \rho_{L,V} =
\pm 1$ and $d\Delta \rho_{L,V}/dt = 0$ at $|t| = 1$.  Near the critical
point, the exponential damping of the first and second terms on the
right hand side of Eq.~(\ref{eq:coex_bkgd}) are negligibly small as
evidenced by the large best fit values $t_1 = 1.56$, $t_2 = 4.55$,
and $b_2 = 0.921$.  As it can be seen in Fig.~\ref{fig:Coex_leveled},
the addition of the saturation background (dashed line) only slightly
improves the systematic difference between the theory and the
experimental data for $|t| < 4\times 10^{-2}$ although it represents the
data quite well for $|t| > 4\times 10^{-2}$.   

\subsection{$\chi_T$, $C_V$, and $|\Delta\rho_{L,V}|$ joint fit}

The good individual fit of the MSR $\phi^4$ model to isothermal
susceptibility $\chi_T$,  specific heat $C_V$, and coexistence
curve $|\Delta\rho_{L,V}|$ has been demonstrated.  A joint fit of all
the three thermal properties leads to a complete test of the MSR
$\phi^4$ model with a minimum set of the parameters, $\{\mu,a, T_c,
C_B\}$.  Here no order parameter saturation was included since its
correction over the fitting range was small.

\begin{figure}
\includegraphics[width=3.1in]{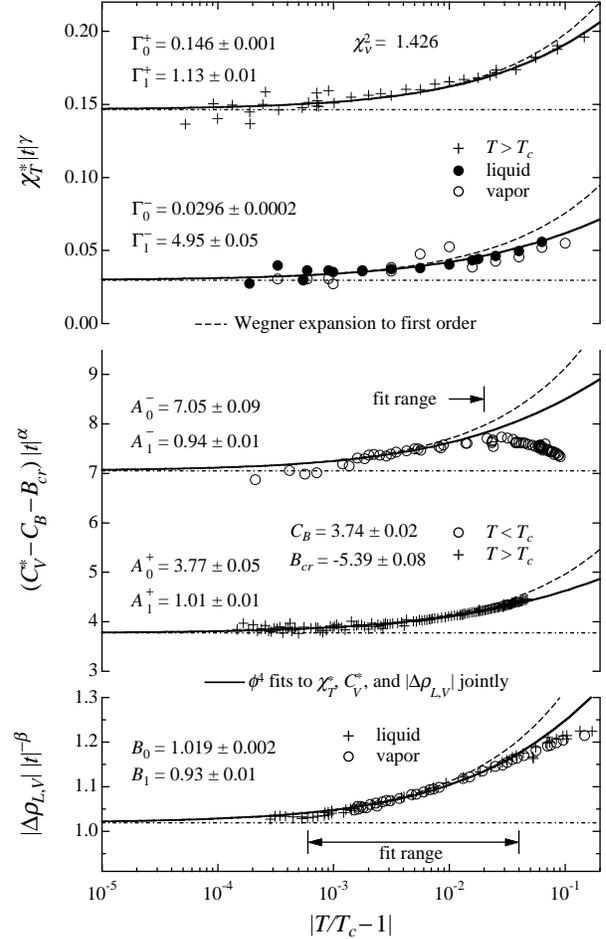}
\caption{A joint fit (solid lines) to susceptibility, specific heat,
and coexistence curve.  The fit used all the shown $\chi_T^*$ data and
the data of $C_V^*$ and $|\Delta\rho_{L,V}|$ over the indicated range. 
The dashed lines are the Wegner expansion to first order with the
listed amplitudes in Table~\protect{\ref{table:joint_fit}}.  The
dot-dashed straight lines represent the asymptotic predictions from the
fit.}
\label{fig:triple_joint_leveled}
\end{figure}

To make sure that no particular measurement dominates the joint fit, a
proper weighting is needed to balance uneven numbers of the experimental
data.   We chose the following weighting in order to normalize the
$\chi^2$ by the number of data points,
\begin{equation}\label{eq:chisqr_joint}
\chi ^2= \frac{N}{3}
\left(\frac{\chi^2_{\chi_T^*}}{N_{\chi_T^*}} +
\frac{\chi^2_{C_V^*}}{N_{C_V^*}} +
\frac{\chi^2_{\Delta\rho_{L,V}}}{N_{\Delta\rho_{L,V}}} \right)\,
\end{equation}
where $N=N_{\chi_T^*}+ N_{C_V^*} + N_{\Delta\rho_{L,V}}$.  In the joint
fit, $\chi_T^*$ and $C_V^*$ were fit against temperature $T$ while
$|\Delta\rho_{L,V}|$ was fit against reduced temperature $|t|$, and
$\mu, a, C_B$, and $T_c$ were adjusted.  The joint fit
results are shown in Fig.~\ref{fig:triple_joint_leveled} and
Table~\ref{table:joint_fit}.   We note that in the joint fit the
uncertainties in $\mu$ and $a$ are much smaller than
in the individual fits even though the overall goodness of fit is worse
in the joint fit.  These improved uncertainties in $\mu$ and $a$
also lead to the improved uncertainties in the critical
amplitudes and the fluctuation-induced background for specific heat.

Shown as dashed lines in Fig.~\ref{fig:triple_joint_leveled} are the Wegner
expansions to first order with the critical amplitudes, $\Gamma_0^\pm$
and $\Gamma_1^\pm$,  $A_0^\pm$ and $A_1^\pm$, $B_0$ and $B_1$,
calculated from the MSR $\phi^4$ model.   Bagnuls and Bervillier
\cite{bagnuls85} have argued that the validity range of any $\phi^4$
model is upper-bounded when the difference between the calculations of
the model and the Wegner expansion to first order becomes significant. 
Based on this argument, the validity range of the MSR $\phi^4$ model is
$|t|\simeq 1\times 10^{-2}$.   However, it is interesting to see that
the MSR $\phi^4$ model provides a good fit beyond $|t| = 1\times
10^{-2}$ to the experimental measurements of the isothermal
susceptibility both above and below $T_c$ and the specific heat above
$T_c$.
\begin{table}
\caption{\label{table:joint_fit}The dimensionless
system-dependent parameters for $^{3}$He.  The adjustable parameters
are obtained from the joint fit of the $\phi^4$ model to the measured
$\chi_T^*$, $C_V^*$, and $|\Delta\rho_{L,V}|$ data of $^3$He.}%
\begin{ruledtabular}
\begin{tabular}{cc}
$T_{c}$ (fit) & $3.315546 \pm 0.000005$\\
$u/u^*$ (fixed) & $0.999$\\
$\mu\times 10^{4}$ (fit) & $1.82\pm 0.02 $\\
$a$ (fit) & $0.132\pm 0.001$ \\
$C_{B}$ (fit) & $3.74\pm 0.02$\\
\hline
$\Gamma_{\rm{0}}^{\rm{+}}$ & $0.146 \pm 0.001$\\
$\Gamma_{\rm{0}}^{\rm{-}}$ & $0.0296 \pm 0.0002$\\
$\Gamma_{\rm{1}}^{\rm{+}}$ & $1.13\pm 0.01$\\
$\Gamma_{\rm{1}}^{\rm{-}}$ & $4.95 \pm 0.05$ \\ 
\hline
$A_{\rm{0}}^{\rm{+}}$ & $3.77 \pm 0.05$\\
$A_{\rm{0}}^{\rm{-}}$ & $7.05 \pm 0.09$\\
$A_{\rm{1}}^{\rm{+}}$ & $1.01 \pm 0.01$\\
$A_{\rm{1}}^{\rm{-}}$ & $0.94 \pm 0.01$\\
$B_{cr}$ & $-5.39 \pm 0.08$\\
\hline
$B_0$ & $1.019 \pm 0.002$\\
$B_1$ & $0.93 \pm 0.01$\\
\hline 
$\xi_{\rm{0}}^{\rm{+}}$ & $0.368 \pm 0.002$\\
$\xi_{\rm{1}}^{\rm{+}}$ & $0.732 \pm 0.007$\\
$\xi_{\rm{0}}^{\rm{-}}$ & $0.259 \pm 0.001$\\
$\xi_{\rm{1}}^{\rm{-}}$ & $0.665 \pm 0.006$\\
\end{tabular}
\end{ruledtabular}
\end{table}

Close to $T_c$, the susceptibility data for $T> T_c$ and the
specific heat data for both $T<T_c$ and $T>T_c$ deviate slightly
from the theoretical prediction.  These deviations can be
attributed to a gravity-induced density stratification.  Since
the specific heat was measured as an average of the whole cell
while the susceptibility was measured locally across a density
sensor, there was a stronger gravity effect in the measured $C_V$ than
$\chi_T$.  The gravity effect on $\chi_T(T<T_c)$ is about a factor of
five smaller than that on $\chi_T(T>T_c)$ because of the difference
in $\chi_T$ magnitudes.  When $T_c$ is used as an adjustable
parameter, the individual fits of susceptibility and specific heat
tend to skew $T_c$ such that the difference between the experimental
measurements and theoretical prediction is minimized because of the
shift in reduced temperature for the measurements.  The $T_c$
determined from the fits of the specific heat
(Fig.~\ref{fig:Cv_leveled}) and susceptibility data
(Fig.~\ref{fig:chi_T_leveled}) tends to be higher and lower, respectively,
than it should be.  This tendency was approximately cancelled out in
the joint fit shown in Fig.~\ref{fig:triple_joint_leveled}.  The slight
gravity effect on the experimental measurements for
$1\times 10^{-4} < |t| < 6\times 10^{-4}$ can be clearly seen in
Fig.~\ref{fig:triple_joint_leveled}.

We mention that in ref.~\cite{meyer01} earlier measurements of the
susceptibility of $^3$He, both above and below $T_c$, were compared with
the present data.  Also in Table I of that reference, the amplitudes of
susceptibility and coexistence curve data and their ratios, such as
$\Gamma_1^+/ B_1$ and $\Gamma_1^+/\Gamma_1^-$, obtained from individual
fits, were presented.

\subsection{Predictions for correlation length and light scattering
intensity}

\begin{figure}
\includegraphics[width=3.1in]{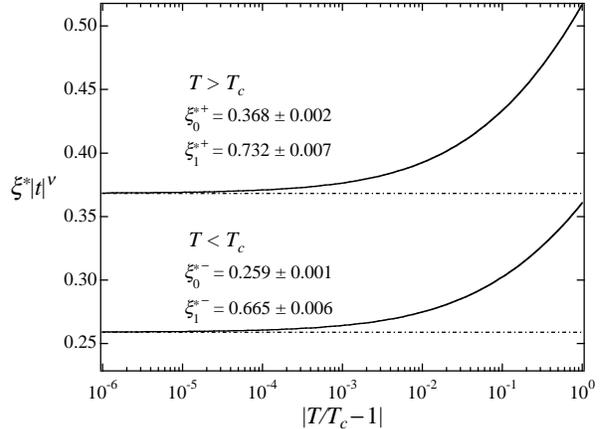}
\caption{The dimensionless correlation length versus reduced
temperature calculated from the MSR $\phi^4$ model for $^3$He.  The
dot-dashed straight lines represent the asymptotic predictions from the
fit.}
\label{fig:xi_leveled}
\end{figure}

By using $u/u^*$, $\mu$, and $a$ given in Table~\ref{table:joint_fit},
the dimensionless correlation length can be calculated for any given
$|t|$ using Eq.~(\ref{eq:correlation_length}).
Figure~\ref{fig:xi_leveled} shows the dimensionless correlation length
versus $t$ calculated from the MSR $\phi^4$ model for $^3$He.  The
length scale to recover the dimensional $\xi_0$ is $l_0$, given by
Eq.~(\ref{eq:l0}).  Thus one has a dimensional $\xi_0 = \xi_0^* l_0 =
2.71\, {\rm\AA}$. This value can be directly compared with $\xi_0 = 2.6
\,{\rm\AA}$ measured in an acoustic experiment ref.~\cite{roe78}. 
Considering that the experimental $\xi_0$ had 10\% uncertainty, the
agreement is very good.

The correlation length can also be determined from a light scattering
experiment.  Miura, Meyer, and Ikushima measured the intensity of
scattered light of $^3$He fluid near its critical point
\cite{miura84}.  The intensity scattered per unit beam length per unit
solid angle in the fluid, $I$, is given by
\begin{equation}
I = I_0 A \chi_T \sin^2 \phi\, g(k \xi) \,,
\end{equation}
where $I_0$ is the beam intensity in the scattering region, $\phi$
is the angle between the electric field of the incident light and
the wave vector of the scattered light, $\chi_T$ is the
susceptibility, and $A = \pi^2 k_B T (\partial n^2 / \partial
\rho)_T^2 / \lambda_0^4$.  Here $n$ is the index of refraction of
the fluid, and $\lambda_0$ is the vacuum wavelength of the
incident light. The function $g(k\xi)$ is, for $k \xi \le 10$,
very accurately given by the Ornstein-Zermike approximation $(1 +
k^2 \xi^2)^{-1+\eta/2}$, where $k$ is the scattering wave vector, $\xi$
is the correlation length, and $\eta$ is the critical exponent of the
fluctuation correlation at the critical point.  In ref.~\cite{miura84},
$k = 5.64\times10^4$~cm$^{-1}$.   At $t = 1\times10^{-6}$, the value of
correlation length can be estimated from $\xi = 2.71 {\rm\AA} t^{-0.63}
= 1.63\times10^{-4}$~cm, hence the condition $k\xi(t = 1\times10^{-6})
= 9.2 \le 10$ was satisfied for $t \ge 1\times 10^{-6}$. Since $B=I_0 A
\sin^2 \phi$ is essentially a constant for the experimental condition,
one can use the knowledge of $\chi_T$ and $\xi$, based on the MSR
$\phi^4$ model, to fit experimental data of the scattered intensity,
with $B$ as an adjustable parameter.  As it can be seen in
Fig.~\ref{fig:light_scattered}, the agreement between the experimental
data and theoretical calculation is reasonably good.  
\begin{figure}
\includegraphics[width=3.1in]{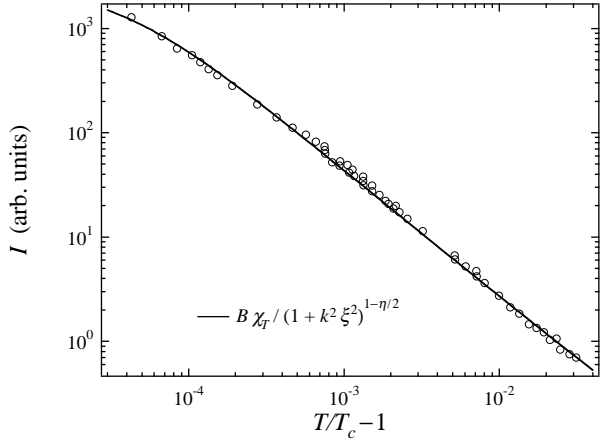}
\caption{\label{fig:light_scattered}The intensity of light scattered by
$^3$He versus reduced temperature.  The theoretical calculated $I$,
using the MSR $\phi^4$ model, is adjusted with a constant amplitude for
$I_0 A$ while $u/u^*$, $\mu$, and $a$ were fixed from the values given
in Table~\protect{\ref{table:joint_fit}}.}
\label{fig:light_scattered}
\end{figure}

\section{Discussion}
In this paper we have used parametric expressions to calculate the
isothermal susceptibility, specific heat, coexistence curve, and
correlation length along the critical isochore and coexistence curve
from the asymptotic region to the crossover region.  All the critical
leading amplitude ratios were contained in the model as listed in
Table~\ref{table:ratios}.   Using only two adjustable parameters in
these theoretical expressions for the critical contributions, we fit
the theory to recently obtained experimental data for the isothermal
susceptibility, specific heat, and early experimental data of the
coexistence curve and light scattering intensity.  The agreement
between the theory and experimental measurements is good.

Further improvements to the minimal renormalization scheme are desired,
especially the five-loop Borel resummations throughout the whole range
of $0 \le u \le u^*$.  More accurate Borel resummations at the fixed
point should also lead to improved calculations of
$\zeta_\phi(u^*)$ and $\zeta_r(u^*)$ so that the resultant critical
exponents can be compared with other published values (see Appendix
\ref{dix:const}).  Theoretical insights on non-critical
contributions are also needed in order to formulate more accurate
analytical expressions for the background contributions.

While the present minimal subtraction renormalization model describes
quite well to experimental measurements along the critical isochore and
coexistence curve, it is not as yet a model for a complete
equation-of-state.  Recently, Agayan {\it et al}. have developed a
phenomenological crossover parametric model (CPM) equation-of-state
that is also based upon RG theory \cite{anisimov01b}.  Within this
model, the internal constants were adjusted such that the critical
leading amplitude ratios agreed with the values in the Fisher and Zinn
column in Table~\ref{table:ratios}.  This CPM model was developed to
fit simple fluids as well as complex fluid systems that exhibit
non-monotonic crossover behavior.  This non-monotonic crossover
behavior could be described by the CPM approach using a finite cut-off
wavelength as an additional fitting parameter.  However, in simple
fluid systems, like $^3$He, crossover behavior of different physical
quantities can be well described within the framework of the field
theoretical $\phi^4$ model without the finite cut-off wavenumber.

NASA supported microgravity flight experiments \cite{MISTE,COEX}
(\url{http://miste.jpl.nasa.gov}), which are under preparation, will
take experimental data of the susceptibility, specific heat, and
coexistence curve in the asymptotic region.  Combining these
microgravity measurements in the asymptotic region with ground-based
measurements in the crossover region should permit a rigorous test of
the predictions of recent renormalization theories.

\begin{acknowledgments}
We are indebted to Dr. R. Haussmann and Prof. J. Rudnick for supporting
the early development of this work and for many stimulating
discussions. We are also grateful to Prof. H. Meyer for a critical
reading of the manuscript and to Dr. M. Weilert for his contribution in
performing the experiments.  The research described in this paper was
carried out at the Jet Propulsion Laboratory, California Institute of
Technology, under contract with the National Aeronautics and Space
Administration.
\end{acknowledgments}

\appendix
\section{The MSR $\phi^4$ model Constants}
\label{dix:const}

The field-theoretic functions, $\zeta_r(u)$, $\zeta_\phi(u)$, and
$\beta_u(u)$, and the amplitude functions, $P_+(u)$, $Q(u)$,
$f_\pm(u)$, $F_\pm(u)$, $A(u)$, and $B(u)$ are known up to
five-loop order from expansions around $u = 0$.  However, these
expansions do not converge.  To overcome this difficulty, these
quantities were expanded to two-loop order and then extrapolation
terms were added to have the functions agree with the calculations of
high-order Borel resummations at the fixed point \cite{schloms89}.  All
these functions have at least one extrapolation term to match the
function's value at the fixed point $u^*$; some functions also have a
second extrapolation term in order to match the value of its derivative
at the fixed point.  Listed in this appendix are the values of these
extrapolation coefficients, their origins, and recent improvements.  
The effects of these coefficient values on the critical exponents and
the fitting quality in this work are discussed.

The extrapolation coefficients for the field-theoretic functions,
$\zeta_r(u)$, $\zeta_\phi(u)$, and $\beta_u(u)$ in
Eqs.~(\ref{eq:zeta_r_0}), (\ref{eq:zeta_phi_0}), (\ref{eq:beta_u_0}) are
$a_1=3075$, $a_2=30390$, $a_3=37.5$, $a_4=14.10$, and $a_5=31.85$.  They
are taken from Table 2 of ref.\cite{schloms89}.

The fixed point value for $u^* = 0.040485$ is solved from the
condition $\beta_u(u^*) = 0$ using the given values for $a_4$ and
$a_5$.  The latest published $u^*$ value for $n = 1$ is \cite{larin98}
\begin{equation}
u^* = 0.0404 \pm 0.0003\,.
\label{eq:ustar}
\end{equation}
The asymptotic critical exponents are linked to the exponent
functions $\zeta_r$, $\zeta_\phi$, and $\beta_u$ by
\begin{equation}
\eta = -\zeta_\phi(u^*) = -\zeta_\phi^* = 0.0367\,,
\label{eq:eta}
\end{equation}
\begin{equation}
\nu = \left[2 - \zeta_r(u^*)\right]^{-1} = \left(2 -
\zeta_r^*\right)^{-1} = 0.629\,,
\label{eq:nu}
\end{equation}
\begin{equation}
\omega = \left.{{d\beta_u(u,
\epsilon=1)}\over{du}}\right|_{u^*} = 0.797\,.
\label{eq:omega}
\end{equation}
Once the critical exponents, $\eta$, $\nu$, and $\omega$ are
known, the remaining important critical exponents can be obtained from
scaling using
\begin{equation}
\alpha = {{1-2\zeta_r^*}\over{2-\zeta_r^*}} = 0.112\,,
\label{eq:alpha}
\end{equation}
\begin{equation}
\beta = {{1-\zeta_\phi^*}\over{2\left(2-\zeta_r^*\right)}} = 0.326\,,
\label{eq:beta}
\end{equation}
\begin{equation}
\gamma = {{2+\zeta_\phi^*}\over{2-\zeta_r^*}} = 1.235\,,
\label{eq:gamma}
\end{equation}
\begin{equation}
\Delta = \nu\omega = 0.502\,.
\label{eq:Wegner_Delta}
\end{equation}
For $n = 1$, the latest theoretically calculated critical
exponents given by Guida and Zinn-Justin \cite{guida98} are,
\begin{equation}
\nu = 0.6304 \pm 0.0013\,,
\label{eq:exponent_nu}
\end{equation}
\begin{equation}
\eta = 0.0335 \pm 0.0025\,,
\label{eq:exponent_eta}
\end{equation}
\begin{equation}
\alpha = 0.109 \pm 0.004\,,
\label{eq:exponent_alpha}
\end{equation}
\begin{equation}
\beta = 0.3258 \pm 0.0014\,,
\label{eq:exponent_beta}
\end{equation}
\begin{equation}
\gamma = 1.2396 \pm 0.0013\,,
\label{eq:exponent_gamma}
\end{equation}
\begin{equation}
\omega = 0.799 \pm 0.011\,,
\label{eq:exponent_omega}
\end{equation}
\begin{equation}
\Delta = \omega\nu = 0.504 \pm 0.008\,,
\label{eq:exponent_Delta}
\end{equation}
A clear difference exits for the value of the critical exponent
$\gamma$ which warrants further efforts from the theoretical
community for improvements in the MSR $\phi^4$ model calculation.

The amplitude function $Q({u}(l))$ for reduced temperature is
expressed as
\begin{equation}\label{eq:Qu}
Q({u}(l)) = 2\int_{u^*}^{{u}(l)} d u^\prime {
{P_+(u^\prime) }\over {\beta_u(u^\prime)}}  \exp
\int_{{u}(l)}^{u^\prime} d u^{\prime\prime} { {2-
\zeta_r(u^{\prime\prime}) }\over {\beta_u(u^{\prime\prime})}}\,.
\end{equation}
At the fixed point $u^*$, there is an identity $\nu^{-1} = 2 -
\zeta_r^*$ that simplifies Eq.~(\ref{eq:Qu}) and leads to
\begin{equation}\label{eq:Qustar}
Q^* = 2\nu P_+^* \,.
\end{equation}
Krause {\em et al}. \cite{krause90} obtained the expression
[Eq.~(K\_A28)]
\begin{equation}\label{eq:dQdu}
\left. {{dQ}\over{du}} \right|_{u^*} = {{2\left. {{dP}\over{du}}
\right|_{u^*} + Q^* \left. {{d\zeta_r}\over{du}}
\right|_{u^*}}\over{\omega + \nu^{-1}}} \,,
\end{equation}
with $\omega = \left. d\beta_u / du \right|_{u^*}$.   They also provided
a one-loop expression for $P(u)$ using a higher-order approximation
[Eq.~(K3.2)]  \cite{krause90}
\begin{equation}
P_+(u) = 1 - 6\, u\, (1 + b_P\, u)\,. \label{eq:Pu1loop}
\end{equation}
The latest calculation by Larin {\em et al}. \cite{larin98} for $n = 1$
gives
\begin{equation}
P_+^* = 0.7568 \pm 0.0044\,.
\label{eq:amplitude_P}
\end{equation}
If the theoretically calculated critical exponent $\nu$ for $n =
1$, given by Guida and Zinn-Justin \cite{guida98}, is used
\cite{larin98}, one has
\begin{eqnarray}
b_+ &=& 2\nu P_+^* = 0.9542 \pm 0.0059, \\  b_- &=& 3/2 - 2\nu P_+^* =
0.5458 \pm  0.0059\,.
\label{eq:b_plus_b_minus2}
\end{eqnarray}
For the extrapolation coefficients in the expression for $Q(u)$,
Eq.~(\ref{eq:Qu1loop}),  Krause {\em et al}. \cite{krause90} determined
$b_Q = 28.2$ and $c_Q = 7.66$ such that Eq.~(\ref{eq:Qustar}) and
(\ref{eq:dQdu}) were satisfied with the then calculated $P_+^*$.  The
values of the extrapolation coefficients $b_Q = 20.32$ and
$c_Q = 6.24$ have been readjusted to agree with the new $Q(u^*) =
0.9542$. There is no value for $\left. dP/du\right|_{u^*}$, so the new
$\left. dQ/du\right|_{u^*}$ has been fixed to its old value
\cite{krause90}.

For the amplitude function $f_\pm(u)$ in the expression of the
susceptibility, Eq.~(\ref{eq:chi2}), $b_\chi = 9.68$ comes from Table 1
of ref.\cite{krause90} and $d_\chi = -11.18$ comes from Table 4 of ref.
\cite{halfkann92}.

For the amplitude function in the expression of the specific heat,
Eqs.~(\ref{eq:C_bare_2}), (\ref{eq:Ku}), and (\ref{eq:amp_cv}), the
five-loop approximation with a Borel resummation gives
\cite{larin98}
\begin{equation}
u^*F_-(u^*) = 0.3687 \pm 0.0040\,.
\label{eq:F_minus}
\end{equation}
By combining Eqs.~(\ref{eq:amp_cv})b, (\ref{eq:ustar}), and
(\ref{eq:F_minus}), the old interpolation coefficient, $d_F = -4.04$
(Table 4 of ref.\cite{halfkann92}), becomes $d_F = -4.6736$. The latest
five-loop calculation also gives \cite{larin98}
\begin{equation}
u^*[F_-(u^*) - F_+(u^*)]= 0.4170 \pm 0.0036\,.
\label{eq:F_difference}
\end{equation}
Using Eqs.~(\ref{eq:F_minus}) and (\ref{eq:F_difference}), one has
\begin{equation}
-u^*F_+(u^*) = 0.0483 \pm 0.0076\,.
\label{eq:F_plus}
\end{equation}
By combining Eqs.~(\ref{eq:amp_cv})a, (\ref{eq:ustar}), and
(\ref{eq:F_plus}), the old interpolation coefficient, $b_F = 5.04$
(Table 1 of ref.\cite{krause90}), becomes $b_F = -5.07$.

We modify Eq.~(\ref{eq:B_u}) to be
\begin{equation}\label{eq:B_u2}
B(u)={\textstyle{1\over 2}} + 9 (1+b_B u) u^2 \,
\end{equation}
with $b_B = -20.68$ in order to satisfy the five-loop Borel resumed
results [Eq.~(L2.34)] \cite{larin98}
\begin{equation}\label{eq:Bstar}
B(u^*) = 0.5024 \pm 0.001\,.
\end{equation}
All the calculations use the value of $u^*$ derived in this paper.

Table~\ref{table:extrapolation} lists the values of the various
extrapolation coefficients for the amplitude functions in the
MSR $\phi^4$ model.  Table~\ref{table:fixedpoint} lists the values of
the various amplitude functions at the fixed point $u^*$. 
\begin{table}
\caption{\label{table:extrapolation}The values of the various
extrapolation coefficients for the amplitude functions in the MSR
$\phi^4$ model.}%
\begin{ruledtabular}
\begin{tabular}
[c]{c|c|c}
coefficient & value  & appeared in\\\hline
$a_1$ & $3075$  & $\zeta_r(u)$ for $Z_r(u)$\\
$a_2$ & $30390$ & $\zeta_r(u)$ for $Z_r(u)$\\
$a_3$ & $37.5$  & $\zeta_\phi(u)$ for $Z_\phi(u)$\\
$a_4$ & $14.10$ & $\beta_u(u)$ for $Z_u(u)\,Z_\phi(u)$\\
$a_5$ & $31.85$ & $\beta_u(u)$ for $Z_u(u)\,Z_\phi(u)$\\
\hline
$b_Q$ & $20.32$ & $Q(u)$ for $t(l)$ \\
$c_Q$ & $6.24$ & $Q(u)$ for $t(l)$\\\hline
$b_\chi$ & $9.68$   & $f_+(u)$ for $\chi_T^+$\\
$d_\chi$ & $-11.18$ & $f_-(u)$ for $\chi_T^-$\\
\hline
$b_F$ & $-5.0726$ & $F_+(u)$ for $C_V^+$\\
$d_F$ & $-4.6736$ & $F_-(u)$ for $C_V^-$\\
\hline 
$b_B$ & $-20.6817$ & $B(u)$ for $C_V^\pm$\\ 
\hline  
$d_\phi$ & $0.702$ & $f_\phi(u)$ for $\Delta\rho_{L,V}$\\
\end{tabular}
\end{ruledtabular}
\end{table}
\begin{table}
\caption{\label{table:fixedpoint}The values of the various
amplitude functions at the fixed point $u^*$.  These values are used in
the calculation of the leading critical amplitude ratios with $u^* =
0.040485$ .}%
\begin{ruledtabular}
\begin{tabular}
[c]{c|c|c}
coefficient & value & appeared in\\\hline
$b_+^*$ & $0.9542 \pm 0.0059$  & $t(l)$ in
Eq.~(\protect{\ref{eq:t2lc}})\\
$b_-^*$ & $0.5458 \pm  0.0059$ & $t(l)$ in
Eq.~(\protect{\ref{eq:t2lc}})\\
\hline
$f_+^*$ & $0.9767$   & $\Gamma_0^+$ for $\chi_T^+$ in
Eq.~(\protect{\ref{eq:Gamma_0}})\\
$f_-^*$ & $2.413$ & $\Gamma_0^-$ for $\chi_T^-$ in
Eq.~(\protect{\ref{eq:Gamma_0}})\\
\hline
$-u^*F_+^*$ & $0.0483 \pm 0.0076$ & $A_0^+$ for $C_\phi^+$ in
Eq.~(\protect{\ref{eq:A_0}})\\
$u^*F_-^*$ & $0.3687 \pm 0.0040$ & $A_0^-$ for $C_\phi^-$ in
Eq.~(\protect{\ref{eq:A_0}})\\
$B^*$ & $0.5024 \pm 0.001$ & $A^*$ for $C_\phi^\pm$ in
Eq.~(\protect{\ref{eq:A_star}})\\ 
\hline  
$f_\phi^*$ & $3.175$ & $B_0$ for $\Delta\rho_{L,V}$ in
Eq.~(\protect{\ref{eq:B_0}})\\
\end{tabular}
\end{ruledtabular}
\end{table}

Equations~(\ref{eq:t2lc}) and (\ref{eq:chi2}) provide a clear
identification of the leading critical divergence and crossover
contribution in a multiplicative form.  In the original expressions,
the critical divergence is contained implicitly in the integrals of
$\zeta_r$ and $\zeta_\phi$ in Eqs.~(\ref{eq:t2la}) and
(\ref{eq:chi1}).  The calculated $\zeta_r(u^*)$ and
$\zeta_\phi(u^*)$ using Borel resummations at the fixed point lead to
the critical exponents $\nu$ and $\eta$ that are slightly different
from the latest values given by Guida and Zinn-Justin.  Because of the
expressions in Eqs.~(\ref{eq:t2lb}) and (\ref{eq:chi2}), the critical
exponent values given by Guida and Zinn-Justin are used for the leading
divergence. The inconsistency is only in the crossover part in the
integrands of $[\zeta_r({u}) -\zeta_r(u^*)]$ and $[\zeta_\phi({u}) -
\zeta_\phi(u^*)]$ that go to zero as the fixed point is approached.

\section{Derivation of Susceptibility Amplitudes}\label{dix:chi_amplitudes}

Expressions for the Wegner expansion of the susceptibility will be
derived in this Appendix that were not presented in previously
published work.  Multiplying Eq.~(\ref{eq:chi2}) by
Eq.~(\ref{eq:t2lc}) to the power $\gamma$ yields
\begin{equation}\label{eq:chi3}
\chi _\pm \, |t|^\gamma = \chi_0 \, \left[b_\pm(l)
t_0\right]^\gamma \, {{\exp [-F_\phi(l) - \gamma F_r(l)]} \over
{f_\pm (l)}}\,.
\end{equation}
In order to expand the exponent functions, $F_r({u}(l))$ and
$F_\phi({u}(l))$, based on Eqs.~(\ref{eq:Fr}) and
(\ref{eq:Fphi}), one needs to expand first the function for the flow
equation, $\beta_u({u}(l))$, to the first order in
$[{u}(l) - u^*]$,
\begin{equation}\label{eq:beta_expanded}
\beta_u({u}(l)) = \omega [{u}(l) - u^*] + O[({u}(l) -
u^*)^2]\,,
\end{equation}
where $\omega = d\beta_u/du|_{u^*}$ and $\beta_u(u^*) = 0$.  Since
$F_r(u^*) = 0$ and $F_\phi(u^*) = 0$,  one obtains
\begin{eqnarray}\label{eq:Fr_expanded}
\nonumber F_r({u}(l)) &=& {\mathop {\lim
}\limits_{{u}(l)\to u^*}} {{\zeta_r({u}(l)) -
\zeta_r(u^*)}\over {\omega [{u}(l) -
u^*]}} \, [{u}(l) - u^*] \\  &+& O[({u}(l) - u^*)^2] \\
\nonumber &=& {{\zeta_r^\prime(u^*)}\over{\omega}}[{u}(l) -
u^*] + O[({u}(l) - u^*)^2]\,,
\end{eqnarray}
\begin{eqnarray}\label{eq:Fphi_expanded}
\nonumber F_\phi({u}(l)) &=& {\mathop {\lim
}\limits_{{u}(l)\to u^*}} {{\zeta_\phi({u}(l)) -
\zeta_\phi(u^*)}\over {\omega [{u}(l) - u^*]}} \, [{u}(l)
- u^*] \\ &+& O[({u}(l) - u^*)^2] \\ \nonumber &=&
{{\zeta_\phi^\prime(u^*)}\over{\omega}}[{u}(l) - u^*] +
O[({u}(l) - u^*)^2]\,,
\end{eqnarray}
\begin{eqnarray}\label{eq:ampl_chi_expanded}
\nonumber f_\pm({u}(l)) &=& f_\pm(u^*)
+ f_\pm^\prime(u^*) [{u}(l) - u^*] \\
 &+& O[({u}(l) - u^*)^2]\,,
\end{eqnarray}
\begin{eqnarray}\label{eq:ampl_t_expanded}
\nonumber b_\pm({u}(l)) &=& b_\pm(u^*)  +
b_\pm^\prime(u^*) [{u}(l) - u^*] \\
 &+& O[({u}(l) - u^*)^2]\,.
\end{eqnarray}
The expression $b_\pm({u}(l))^\gamma\exp[-F_\phi({u}(l))-\gamma
F_r({u}(l))]  /  f_\pm ({u}(l))$ is then expanded in terms of $[{u}(l) -
u^*]$, dropping the higher orders, to give
{\setlength{\arraycolsep}{0pt}
\begin{eqnarray}\label{eq:chi4}
&&\chi _\pm \, |t|^\gamma = {{\chi_0 \, (b_\pm^* t_0)^\gamma}\over
{f_\pm(u^*)}}  \times \\ \nonumber && \left\{1 - \left.\left[
\gamma\left( {{\zeta_r^\prime}\over{\omega}} -
{{b_\pm^\prime}\over{b_\pm}} \right) +
{{\zeta_\phi^\prime}\over{\omega}} + {{f_\pm^\prime}\over{f_\pm}}
\right]\right|_{u^*} \, [{u}(l) - u^*] \right\}\,.\nonumber
\end{eqnarray}}
The solution of the flow equation with $\beta_u(u)$ approximated
by Eq.~(\ref{eq:beta_expanded}) is
\begin{equation}\label{eq:l_expanded}
l^\omega = {{{u}(l) - u^*}\over {u - u^*}}\,.
\end{equation}
By expressing $l$ in terms of $|t|$ and dropping higher order
terms, one has $l = |t|^\nu/(b_\pm^* t_0)^\nu$ and
{\setlength{\arraycolsep}{1pt}
\begin{eqnarray}\label{eq:chi5}
\chi _\pm \, |t|^\gamma &=& {{\chi_0 \, (b_\pm^* t_0)^\gamma}\over
{f_\pm(u^*)}} \, \\ \nonumber &\times& \left[1 - \left.\left(
\gamma{{\zeta_r^\prime}\over{\omega}} - \gamma
{{b_\pm^\prime}\over{b_\pm}} + {{\zeta_\phi^\prime}\over{\omega}}
+ {{f_\pm^\prime}\over{f_\pm}} \right)\right|_{u^*} \right. \\
\nonumber & & \,\,\,\,\,\,\,\,\,\times \left. {{u - u^*}\over
{(b_\pm^* t_0)^{\nu\omega}}} |t|^{\nu\omega}\right]\,.
\end{eqnarray}}
Comparing Eq.~(\ref{eq:chi5}) to the standard Wegner expansion to
the first term, see Eq.~(\ref{eq:chi_Wegner}), one obtains the
critical amplitudes of the susceptibility expressed analytically
in Eqs.~(\ref{eq:Gamma_0}) and (\ref{eq:Gamma_1}) with $\Delta =
\nu\omega$.

\section{Derivation of Specific Heat Amplitudes}\label{dix:cv_amplitudes}

A derivation of the critical amplitudes and constant background of
the specific heat in the additive renormalization form will be given in
this Appendix.   This derivation is consistent with the one for
susceptibility given above and is different from the one given by
Schloms and Dohm \cite{schloms90}.

First an expansion expression for the function $A(u(l))$ will
be derived that is an approximate solution of
Eq.~(\ref{eq:Au_diff}).  By expanding $B(u)$ and $\zeta_r(u)$ around
$u^*$ and omitting higher order terms beyond the linear term,
Eq.~(\ref{eq:Au_diff}) becomes
{\setlength{\arraycolsep}{1pt}
\begin{eqnarray}\label{eq:Au_diff_approx_1}
l{{dA (l)} \over {dl}}&=& 4B(u^*) +4B^\prime(u^*)({u} - u^*)
\\ \nonumber &+& \left[\alpha/\nu - 2\zeta _r^\prime(u^*) (
{u} - u^*) \right] A(l)\,.
\end{eqnarray}}
Then Eq.~(\ref{eq:l_expanded}) is used to replace $({u} -
u^*)$ with $l^\omega$, yielding
\begin{equation}\label{eq:Au_diff_approx_2}
l{{dA(l)} \over {dl}}= H + Y l^\omega + \left(G + Z l^\omega
 \right) A(l)\,.
\end{equation}
where
{\setlength{\arraycolsep}{1pt}
\begin{eqnarray}
H &=& 4B(u^*) \label{eq:Au_sub_H}\\
G &=& \frac{\alpha}{\nu} \label{eq:Au_sub_G}\\
Y &=& 4B^\prime(u^*)(u - u^*) \label{eq:Au_sub_Y}\\
Z &=& - 2\zeta _r^\prime(u^*) ( u - u^*)\,. \label{eq:Au_sub_Z}
\end{eqnarray}}
With a variable change of
\begin{equation}
v = \frac{l^{\omega }\,Z}{\omega} \label{eq:lomegaZ}\,,
\end{equation}
Eq.~(\ref{eq:Au_diff_approx_2}) becomes
\begin{equation}\label{eq:Au_diff_approx_3}
v{{dA(v)} \over {dv}}= \frac{H}{\omega} + \frac{Y}{Z} v +
\left(\frac{G}{\omega} + v \right) A(v)\,.
\end{equation}
The solution of Eq.~(\ref{eq:Au_diff_approx_3}) is
{\setlength{\arraycolsep}{1pt}
\begin{eqnarray}\label{eq:Au_solution_1}
A(v) &=& \exp(v)\,v^{\frac{G}{\omega }}\\ \nonumber &\times&
\left[K_1 - \frac{H}{\omega} \Gamma\left(-\frac{G}{\omega },
v\right) - \frac{Y}{Z} \Gamma\left(1-\frac{G}{\omega },v\right)
\right]\,,
\end{eqnarray}}
where $K_1$ is a constant to be determined through the initial
condition.  Expanding Eq.~(\ref{eq:Au_solution_1}) in $v$ and
keeping only the linear terms of $l^{\alpha/\nu}$ and $l^\omega$,
one obtains
\begin{equation}\label{eq:Au_solution_2}
A(l) = A(u^*) + K_2(u - u^*)\,l^\omega +  K_3\,l^{\alpha /
\nu}\,,
\end{equation}
where Eq.~(\ref{eq:A_star}) is used for $A(u^*)$ and
{\setlength{\arraycolsep}{1pt}
\begin{eqnarray}\label{eq:K2}
K_2 &=& \frac{1}{(u - u^*)}\left[\frac{\nu Y}{\Delta - \alpha} +
A^*\frac{\Delta}{\Delta -\alpha} \frac{Z}{\omega}\right]
\\\nonumber &=& \frac{2\nu}{\Delta - \alpha}\left[2B^\prime(u^*)
- A^*\zeta _r^\prime(u^*)\right]\,.
\end{eqnarray}}
$K_3$ in Eq.~(\ref{eq:Au_solution_2}) will be eliminated through
initial condition at the reference point $l=1$
\begin{equation}\label{eq:K3a}
K_3 = A_1 - A^* - K_2(u - u^*)\,,
\end{equation}
where $A_1 \equiv A(l=1)$.  Since $A_1$ has not been given as a fitting
parameter, it is calculated from the numerical solution of
Eq.~(\ref{eq:Au_diff}) with $A(u=0) = -2$.

Substituting Eq.~(\ref{eq:Au_solution_2}) into
Eq.~(\ref{eq:C_bare_2}), one has
{\setlength{\arraycolsep}{1pt}
\begin{eqnarray}\label{eq:C_bare_4}
\nonumber \frac{C_\phi^\pm }{C_0} &=&  -\exp [2 F_r(l)] K_3
+ \exp [2 F_r(l)] l^{-\alpha/\nu}\\
 &\times&  \left[ F_\pm(l) - A^* -
K_2 (u - u^*)l^\omega \right]\,.
\end{eqnarray}}
Replacing $l^{\alpha / \nu}$ in Eq.~(\ref{eq:C_bare_4}) with
$t/t_0$ from Eq.~(\ref{eq:t2lc}) leads to
{\setlength{\arraycolsep}{1pt}
\begin{eqnarray}\label{eq:C_bare_5}
 \frac{C_\phi^\pm }{C_0}  &=& -\exp [2 F_r(l)]K_3
\\ \nonumber &+&  \exp [(2-\alpha) F_r(l)] \left[b_\pm(l) t_0\right]^\alpha
|t|^{-\alpha} \\\nonumber  &\times& \left[ F_\pm(l) - A^* - K_2 (u -
u^*)l^\omega
\right]\,.
\end{eqnarray}}
$F_r(l)$, $F_\pm(l)$, and $b_\pm(l)$ are expanded according to
Eqs.~(\ref{eq:Fr_expanded}), (\ref{eq:ampl_chi_expanded}), and
(\ref{eq:ampl_t_expanded}) respectively.  Higher order terms than
$O[({u}(l) - u^*)^2]$ or $O(l^{2\omega})$ are dropped in the
expansion, and $({u}(l) - u^*)$ is replaced using
Eq.~(\ref{eq:l_expanded}).  By using the approximation of
$l^\omega = |t|^\Delta/(b_\pm^* t_0)^\Delta$,  one finally has
{\setlength{\arraycolsep}{1pt}
\begin{eqnarray}\label{eq:C_bare_expanded}
\nonumber\frac{C_\phi^\pm }{C_0} &=& - K_3 + \left( F_\pm^*
-A^*\right) \left(b_\pm^* t_0\right)^\alpha |t|^{-\alpha} \\
\nonumber &\times& \left\{1 + \left[\frac{1}{F_\pm^* -A^*} \left(
F_\pm^\prime - K_2 \right) \right.\right.
\\ \nonumber  &+& \left. \left.(2-\alpha)\frac{\zeta_r^\prime}{\omega}
  + \alpha \frac{b_\pm^\prime}{b_\pm} \right]\right|_{u^*} \\
&\times& \left. (u-u^*){{|t|^\Delta}\over{\left(b_\pm^* t_0\right)^\Delta}}
\right\} \,.
\end{eqnarray}}
By comparing Eq.~(\ref{eq:C_bare_expanded}) to the standard Wegner
expansion to the first term, see Eq.~(\ref{eq:cv_Wegner}), one obtains
the analytical expressions for the critical amplitudes of the specific
heat given in Eqs.~(\ref{eq:A_0}) and (\ref{eq:A_1}).  The critical
background specific heat is also identified as
\begin{equation}\label{eq:Bcr}
B_{cr} = -C_0 K_3\,.
\end{equation}

\section{$\chi_T^*$ and $C_V^*$ experimental measurements}
\label{dix:data}

We list in this Appendix the dimensionless experimental measurements of
isothermal susceptibility, specific heat, and coexistence curve of
$^3$He. The ITS90 temperature standard was used in the following tables.
\begin{longtable}[c]{cccc}
3.3157200 & 5.288e-05 & 2.763e+04 & 0 \kill
\caption{\label{table:expt_data_chi_T}The dimensionless experimental
measurements of the $^3$He isothermal susceptibility $\chi_T^*$.  $T_c
= 3.315545$K was obtained from the joint fit of $\chi_T^*$ and $C_V^*$
to the MSR$\phi^4$ model.  The index = 0, 19, 8 corresponds
respectively to $T>T_c$,$\{T<T_c, {\rm liquid}\}$, $\{T<T_c, {\rm
vapor}\}$.}\\
\endfirsthead
$T$ & $T/T_c - 1$ & $\chi_T^*$ & index \\
\hline
3.3157200 & 5.288e-05 & 2.763e+04 & 0 \\
3.3158500 & 9.209e-05 & 1.524e+04 & 0 \\
3.3158770 & 1.002e-04 & 1.278e+04 & 0 \\
3.3159900 & 1.343e-04 & 9.468e+03 & 0 \\
3.3161700 & 1.886e-04 & 6.006e+03 & 0 \\
3.3161770 & 1.907e-04 & 5.600e+03 & 0 \\
3.3163600 & 2.459e-04 & 4.477e+03 & 0 \\
3.3164000 & 2.580e-04 & 4.457e+03 & 0 \\
3.3166570 & 3.355e-04 & 2.966e+03 & 0 \\
3.3173270 & 5.376e-04 & 1.670e+03 & 0 \\
3.3175600 & 6.078e-04 & 1.469e+03 & 0 \\
3.3179100 & 7.134e-04 & 1.256e+03 & 0 \\
3.3179170 & 7.155e-04 & 1.180e+03 & 0 \\
3.3179400 & 7.225e-04 & 1.194e+03 & 0 \\
3.3180170 & 7.457e-04 & 1.120e+03 & 0 \\
3.3185400 & 9.034e-04 & 9.472e+02 & 0 \\
3.3188970 & 1.011e-03 & 7.790e+02 & 0 \\
3.3202400 & 1.416e-03 & 5.268e+02 & 0 \\
3.3221970 & 2.006e-03 & 3.450e+02 & 0 \\
3.3239900 & 2.547e-03 & 2.587e+02 & 0 \\
3.3259870 & 3.149e-03 & 1.970e+02 & 0 \\
3.3295500 & 4.224e-03 & 1.407e+02 & 0 \\
3.3341869 & 5.623e-03 & 9.850e+01 & 0 \\
3.3404500 & 7.512e-03 & 7.046e+01 & 0 \\
3.3487469 & 1.001e-02 & 4.980e+01 & 0 \\
3.3487600 & 1.002e-02 & 4.979e+01 & 0 \\
3.3596699 & 1.331e-02 & 3.535e+01 & 0 \\
3.3745467 & 1.780e-02 & 2.480e+01 & 0 \\
3.3897402 & 2.238e-02 & 1.884e+01 & 0 \\
3.3999300 & 2.545e-02 & 1.642e+01 & 0 \\
3.3999967 & 2.547e-02 & 1.620e+01 & 0 \\
3.4402800 & 3.762e-02 & 1.014e+01 & 0 \\
3.4999362 & 5.561e-02 & 6.540e+00 & 0 \\
3.4999899 & 5.563e-02 & 6.540e+00 & 0 \\
3.6000160 & 8.580e-02 & 3.990e+00 & 0 \\
3.6000502 & 8.581e-02 & 3.939e+00 & 0 \\
3.8000406 & 1.461e-01 & 2.126e+00 & 0 \\
3.3149265 & -1.864e-04 & 1.151e+03 & 19 \\
3.3144595 & -3.273e-04 & 8.291e+02 & 19 \\
3.3137295 & -5.475e-04 & 3.271e+02 & 19 \\
3.3135883 & -5.901e-04 & 3.665e+02 & 19 \\
3.3125410 & -9.059e-04 & 2.145e+02 & 19 \\
3.3122627 & -9.899e-04 & 1.868e+02 & 19 \\
3.3096911 & -1.765e-03 & 9.418e+01 & 19 \\
3.3050875 & -3.154e-03 & 4.708e+01 & 19 \\
3.2969714 & -5.602e-03 & 2.344e+01 & 19 \\
3.2824717 & -9.975e-03 & 1.225e+01 & 19 \\
3.2630551 & -1.583e-02 & 7.408e+00 & 19 \\
3.2566732 & -1.776e-02 & 6.555e+00 & 19 \\
3.2323114 & -2.510e-02 & 4.455e+00 & 19 \\
3.1836186 & -3.979e-02 & 2.696e+00 & 19 \\
3.1064086 & -6.308e-02 & 1.717e+00 & 19 \\
3.3144595 & -3.273e-04 & 6.347e+02 & 8 \\
3.3137295 & -5.475e-04 & 3.326e+02 & 8 \\
3.3135883 & -5.901e-04 & 3.080e+02 & 8 \\
3.3125410 & -9.059e-04 & 1.820e+02 & 8 \\
3.3122627 & -9.899e-04 & 1.436e+02 & 8 \\
3.3096911 & -1.765e-03 & 9.300e+01 & 8 \\
3.3051276 & -3.142e-03 & 4.865e+01 & 8 \\
3.3050875 & -3.154e-03 & 4.533e+01 & 8 \\
3.2969591 & -5.606e-03 & 2.937e+01 & 8 \\
3.2824590 & -9.979e-03 & 1.586e+01 & 8 \\
3.2630538 & -1.583e-02 & 6.585e+00 & 8 \\
3.2323290 & -2.510e-02 & 4.100e+00 & 8 \\
3.1836229 & -3.979e-02 & 2.486e+00 & 8 \\
3.1064148 & -6.308e-02 & 1.600e+00 & 8 \\
2.9840784 & -9.997e-02 & 9.556e-01 & 8 \\
\hline \hline
\end{longtable}

\begin{longtable}[c]{ccc}
3.3157200 & 5.288e-05 & 2.763e+04\kill
\caption{The dimensionless experimental measurements of the $^3$He
specific heat $C_V^*$.  $T_c = 3.315545$K was obtained
from the joint fit of $\chi_T^*$ and $C_V^*$ to the MSR $\phi^4$
model.\label{table:expt_data_cv}}%
\endfirsthead
$T$ & $T/T_c - 1$ & $C_V^*$\\\hline
3.014570 & -9.078e-02 & 7.881 \\
3.021686 & -8.863e-02 & 7.946 \\
3.028699 & -8.652e-02 & 7.959 \\
3.035582 & -8.444e-02 & 8.038 \\
3.042383 & -8.239e-02 & 8.088 \\
3.049135 & -8.035e-02 & 8.069 \\
3.055774 & -7.835e-02 & 8.173 \\
3.059940 & -7.709e-02 & 8.183 \\
3.064174 & -7.582e-02 & 8.200 \\
3.070900 & -7.379e-02 & 8.231 \\
3.077578 & -7.177e-02 & 8.312 \\
3.081723 & -7.052e-02 & 8.331 \\
3.085888 & -6.927e-02 & 8.332 \\
3.092441 & -6.729e-02 & 8.541 \\
3.099071 & -6.529e-02 & 8.471 \\
3.103159 & -6.406e-02 & 8.428 \\
3.104792 & -6.357e-02 & 8.498 \\
3.106435 & -6.307e-02 & 8.512 \\
3.108071 & -6.258e-02 & 8.498 \\
3.109702 & -6.208e-02 & 8.543 \\
3.111329 & -6.159e-02 & 8.578 \\
3.112949 & -6.110e-02 & 8.624 \\
3.114572 & -6.062e-02 & 8.528 \\
3.115721 & -6.027e-02 & 8.561 \\
3.116924 & -5.991e-02 & 8.622 \\
3.120951 & -5.869e-02 & 8.631 \\
3.130580 & -5.579e-02 & 8.680 \\
3.143237 & -5.197e-02 & 8.798 \\
3.155746 & -4.820e-02 & 8.902 \\
3.165016 & -4.540e-02 & 9.008 \\
3.171098 & -4.357e-02 & 9.061 \\
3.177146 & -4.174e-02 & 9.118 \\
3.182946 & -3.999e-02 & 9.177 \\
3.188913 & -3.819e-02 & 9.248 \\
3.190517 & -3.771e-02 & 9.231 \\
3.191638 & -3.737e-02 & 9.257 \\
3.198331 & -3.535e-02 & 9.419 \\
3.210298 & -3.174e-02 & 9.590 \\
3.222044 & -2.820e-02 & 9.760 \\
3.235858 & -2.403e-02 & 9.677 \\
3.237019 & -2.368e-02 & 9.888 \\
3.237827 & -2.344e-02 & 9.861 \\
3.238887 & -2.312e-02 & 9.836 \\
3.245184 & -2.122e-02 & 10.083 \\
3.268550 & -1.417e-02 & 10.452 \\
3.269553 & -1.387e-02 & 10.481 \\
3.286051 & -8.896e-03 & 11.023 \\
3.287624 & -8.421e-03 & 10.997 \\
3.288580 & -8.133e-03 & 11.091 \\
3.296766 & -5.664e-03 & 11.477 \\
3.297686 & -5.386e-03 & 11.591 \\
3.298910 & -5.017e-03 & 11.697 \\
3.301335 & -4.286e-03 & 11.856 \\
3.304018 & -3.477e-03 & 12.119 \\
3.305181 & -3.126e-03 & 12.209 \\
3.306067 & -2.859e-03 & 12.272 \\
3.307219 & -2.511e-03 & 12.498 \\
3.308159 & -2.228e-03 & 12.715 \\
3.308833 & -2.024e-03 & 12.833 \\
3.309386 & -1.858e-03 & 12.824 \\
3.310214 & -1.608e-03 & 13.081 \\
3.311035 & -1.360e-03 & 13.029 \\
3.311577 & -1.197e-03 & 13.314 \\
3.313179 & -7.135e-04 & 13.795 \\
3.313675 & -5.639e-04 & 14.128 \\
3.314163 & -4.167e-04 & 14.827 \\
3.314846 & -2.107e-04 & 15.637 \\
3.316036 & 1.482e-04 & 8.389 \\
3.316098 & 1.669e-04 & 8.595 \\
3.316160 & 1.856e-04 & 8.159 \\
3.316217 & 2.028e-04 & 8.289 \\
3.316279 & 2.215e-04 & 7.875 \\
3.316339 & 2.396e-04 & 7.907 \\
3.316400 & 2.580e-04 & 7.823 \\
3.316405 & 2.595e-04 & 7.992 \\
3.316461 & 2.764e-04 & 7.654 \\
3.316512 & 2.918e-04 & 7.572 \\
3.316562 & 3.068e-04 & 7.689 \\
3.316627 & 3.264e-04 & 7.617 \\
3.316690 & 3.454e-04 & 7.643 \\
3.316757 & 3.656e-04 & 7.274 \\
3.316827 & 3.868e-04 & 7.473 \\
3.316893 & 4.067e-04 & 7.407 \\
3.316943 & 4.217e-04 & 7.611 \\
3.316995 & 4.374e-04 & 7.202 \\
3.317088 & 4.655e-04 & 7.302 \\
3.317206 & 5.011e-04 & 6.963 \\
3.317399 & 5.593e-04 & 7.138 \\
3.317663 & 6.389e-04 & 6.968 \\
3.317885 & 7.059e-04 & 6.913 \\
3.317972 & 7.321e-04 & 6.858 \\
3.318097 & 7.698e-04 & 6.693 \\
3.318225 & 8.084e-04 & 6.562 \\
3.318429 & 8.699e-04 & 6.581 \\
3.318670 & 9.426e-04 & 6.702 \\
3.318830 & 9.909e-04 & 6.573 \\
3.318919 & 1.018e-03 & 6.624 \\
3.319077 & 1.065e-03 & 6.661 \\
3.319217 & 1.108e-03 & 6.493 \\
3.319447 & 1.177e-03 & 6.449 \\
3.319738 & 1.265e-03 & 6.279 \\
3.320283 & 1.429e-03 & 6.528 \\
3.320536 & 1.505e-03 & 6.285 \\
3.320809 & 1.588e-03 & 6.270 \\
3.320989 & 1.642e-03 & 6.261 \\
3.321325 & 1.743e-03 & 6.209 \\
3.321702 & 1.857e-03 & 6.096 \\
3.322174 & 1.999e-03 & 5.919 \\
3.322638 & 2.139e-03 & 6.104 \\
3.323109 & 2.281e-03 & 5.910 \\
3.323586 & 2.425e-03 & 5.944 \\
3.324054 & 2.566e-03 & 5.853 \\
3.324387 & 2.667e-03 & 5.711 \\
3.324605 & 2.733e-03 & 5.867 \\
3.324881 & 2.816e-03 & 5.874 \\
3.325280 & 2.936e-03 & 5.802 \\
3.325758 & 3.080e-03 & 5.776 \\
3.326317 & 3.249e-03 & 5.741 \\
3.327445 & 3.589e-03 & 5.682 \\
3.328419 & 3.883e-03 & 5.649 \\
3.329076 & 4.081e-03 & 5.485 \\
3.329729 & 4.278e-03 & 5.666 \\
3.331770 & 4.894e-03 & 5.507 \\
3.332809 & 5.207e-03 & 5.440 \\
3.333808 & 5.508e-03 & 5.465 \\
3.334811 & 5.811e-03 & 5.397 \\
3.335822 & 6.116e-03 & 5.382 \\
3.336837 & 6.422e-03 & 5.337 \\
3.337850 & 6.727e-03 & 5.329 \\
3.338592 & 6.951e-03 & 5.308 \\
3.339364 & 7.184e-03 & 5.301 \\
3.340384 & 7.492e-03 & 5.277 \\
3.341933 & 7.959e-03 & 5.241 \\
3.344007 & 8.585e-03 & 5.199 \\
3.346105 & 9.217e-03 & 5.161 \\
3.348203 & 9.850e-03 & 5.142 \\
3.350301 & 1.048e-02 & 5.112 \\
3.352403 & 1.112e-02 & 5.085 \\
3.354507 & 1.175e-02 & 5.051 \\
3.356612 & 1.239e-02 & 5.022 \\
3.357923 & 1.278e-02 & 5.015 \\
3.358726 & 1.302e-02 & 4.987 \\
3.359605 & 1.329e-02 & 5.006 \\
3.361045 & 1.372e-02 & 4.992 \\
3.363198 & 1.437e-02 & 4.967 \\
3.365356 & 1.502e-02 & 4.949 \\
3.367521 & 1.568e-02 & 4.931 \\
3.369688 & 1.633e-02 & 4.909 \\
3.371862 & 1.699e-02 & 4.884 \\
3.374039 & 1.764e-02 & 4.875 \\
3.375402 & 1.805e-02 & 4.917 \\
3.376038 & 1.825e-02 & 4.857 \\
3.376954 & 1.852e-02 & 4.862 \\
3.378597 & 1.902e-02 & 4.845 \\
3.380724 & 1.966e-02 & 4.836 \\
3.382861 & 2.030e-02 & 4.820 \\
3.385006 & 2.095e-02 & 4.812 \\
3.387156 & 2.160e-02 & 4.803 \\
3.389310 & 2.225e-02 & 4.789 \\
3.391472 & 2.290e-02 & 4.774 \\
3.393640 & 2.355e-02 & 4.766 \\
3.395811 & 2.421e-02 & 4.765 \\
3.397990 & 2.487e-02 & 4.738 \\
3.400172 & 2.552e-02 & 4.736 \\
3.402356 & 2.618e-02 & 4.727 \\
3.404543 & 2.684e-02 & 4.720 \\
3.406731 & 2.750e-02 & 4.718 \\
3.408924 & 2.816e-02 & 4.710 \\
3.411116 & 2.883e-02 & 4.701 \\
3.412433 & 2.922e-02 & 4.666 \\
3.413044 & 2.941e-02 & 4.661 \\
3.413649 & 2.959e-02 & 4.737 \\
3.414220 & 2.976e-02 & 4.723 \\
3.414872 & 2.996e-02 & 4.681 \\
3.415805 & 3.024e-02 & 4.678 \\
3.417482 & 3.075e-02 & 4.674 \\
3.419719 & 3.142e-02 & 4.661 \\
3.421956 & 3.209e-02 & 4.660 \\
3.424190 & 3.277e-02 & 4.651 \\
3.426428 & 3.344e-02 & 4.645 \\
3.428664 & 3.412e-02 & 4.631 \\
3.430897 & 3.479e-02 & 4.630 \\
3.433131 & 3.547e-02 & 4.632 \\
3.435362 & 3.614e-02 & 4.647 \\
3.437601 & 3.681e-02 & 4.628 \\
3.439850 & 3.749e-02 & 4.626 \\
3.442101 & 3.817e-02 & 4.620 \\
3.444355 & 3.885e-02 & 4.609 \\
3.446613 & 3.953e-02 & 4.597 \\
3.448872 & 4.021e-02 & 4.592 \\
3.451131 & 4.089e-02 & 4.594 \\
3.453395 & 4.158e-02 & 4.584 \\
3.455659 & 4.226e-02 & 4.576 \\
3.459058 & 4.328e-02 & 4.579 \\
3.463590 & 4.465e-02 & 4.570 \\
3.468123 & 4.602e-02 & 4.557 \\
3.472658 & 4.739e-02 & 4.562 \\\hline \hline
\end{longtable}

\begin{longtable}[c]{ccc}
3.3157200 & 5.288e-05 & 2.763e+04\kill
\caption{The experimental measurements of the $^3$He
reduced density $\Delta\rho_{L,V}$ along the coexistence curve.  The
data was provided by Prof. H. Meyer as it was used in
ref.~\protect{\cite{luijten02}}.  The index = 0, 8 corresponds to
liquid and vapor respectively.
\label{table:expt_data_coex}}%
\endfirsthead
$T/T_c - 1$ & $|\Delta\rho_{L,V}|$ & index\\\hline
2.848e-04 & 7.192e-02 & 0 \\
3.128e-04 & 7.460e-02 & 0 \\
3.445e-04 & 7.705e-02 & 0 \\
3.836e-04 & 7.972e-02 & 0 \\
4.460e-04 & 8.365e-02 & 0 \\
5.319e-04 & 8.859e-02 & 0 \\
5.322e-04 & 8.826e-02 & 0 \\
5.963e-04 & 9.177e-02 & 0 \\
6.638e-04 & 9.512e-02 & 0 \\
7.279e-04 & 9.822e-02 & 0 \\
8.332e-04 & 1.028e-01 & 0 \\
9.630e-04 & 1.089e-01 & 0 \\
9.658e-04 & 1.084e-01 & 0 \\
1.279e-03 & 1.198e-01 & 0 \\
1.399e-03 & 1.235e-01 & 0 \\
1.430e-03 & 1.239e-01 & 0 \\
1.518e-03 & 1.270e-01 & 0 \\
2.171e-03 & 1.438e-01 & 0 \\
2.421e-03 & 1.494e-01 & 0 \\
3.255e-03 & 1.657e-01 & 0 \\
3.862e-03 & 1.761e-01 & 0 \\
4.066e-03 & 1.793e-01 & 0 \\
4.868e-03 & 1.908e-01 & 0 \\
6.266e-03 & 2.090e-01 & 0 \\
7.261e-03 & 2.207e-01 & 0 \\
9.366e-03 & 2.420e-01 & 0 \\
1.319e-02 & 2.738e-01 & 0 \\
1.834e-02 & 3.081e-01 & 0 \\
2.285e-02 & 3.340e-01 & 0 \\
2.854e-02 & 3.626e-01 & 0 \\
3.135e-02 & 3.755e-01 & 0 \\
3.496e-02 & 3.911e-01 & 0 \\
3.783e-02 & 4.027e-01 & 0 \\
4.348e-02 & 4.232e-01 & 0 \\
5.230e-02 & 4.463e-01 & 0 \\
5.320e-02 & 4.474e-01 & 0 \\
5.559e-02 & 4.636e-01 & 0 \\
6.988e-02 & 5.038e-01 & 0 \\
7.140e-02 & 5.124e-01 & 0 \\
7.389e-02 & 5.135e-01 & 0 \\
8.380e-02 & 5.381e-01 & 0 \\
9.589e-02 & 5.656e-01 & 0 \\
1.025e-01 & 5.749e-01 & 0 \\
1.034e-01 & 5.800e-01 & 0 \\
1.339e-01 & 6.363e-01 & 0 \\
1.710e-01 & 6.887e-01 & 0 \\
2.227e-01 & 7.524e-01 & 0 \\
3.106e-01 & 8.389e-01 & 0 \\
4.970e-01 & 9.499e-01 & 0 \\
5.850e-01 & 9.813e-01 & 0 \\
6.980e-01 & 1.007e+00 & 0 \\
7.580e-01 & 1.016e+00 & 0 \\
8.190e-01 & 1.018e+00 & 0 \\
8.790e-01 & 1.019e+00 & 0 \\
9.390e-01 & 1.019e+00 & 0 \\
9.980e-01 & 1.017e+00 & 0 \\
8.433e-04 & 1.039e-01 & 0 \\
8.995e-04 & 1.062e-01 & 0 \\
9.060e-04 & 1.058e-01 & 0 \\
1.221e-03 & 1.172e-01 & 0 \\
1.515e-03 & 1.264e-01 & 8 \\
1.611e-03 & 1.300e-01 & 8 \\
1.647e-03 & 1.302e-01 & 8 \\
1.729e-03 & 1.330e-01 & 8 \\
1.791e-03 & 1.342e-01 & 8 \\
1.931e-03 & 1.375e-01 & 8 \\
2.075e-03 & 1.417e-01 & 8 \\
2.315e-03 & 1.473e-01 & 8 \\
2.382e-03 & 1.478e-01 & 8 \\
2.661e-03 & 1.545e-01 & 8 \\
2.849e-03 & 1.585e-01 & 8 \\
3.085e-03 & 1.625e-01 & 8 \\
3.693e-03 & 1.731e-01 & 8 \\
4.713e-03 & 1.890e-01 & 8 \\
5.169e-03 & 1.950e-01 & 8 \\
6.004e-03 & 2.057e-01 & 8 \\
6.811e-03 & 2.153e-01 & 8 \\
7.311e-03 & 2.205e-01 & 8 \\
8.153e-03 & 2.294e-01 & 8 \\
1.314e-02 & 2.727e-01 & 8 \\
1.500e-02 & 2.857e-01 & 8 \\
1.934e-02 & 3.140e-01 & 8 \\
2.485e-02 & 3.438e-01 & 8 \\
2.880e-02 & 3.617e-01 & 8 \\
3.157e-02 & 3.747e-01 & 8 \\
3.505e-02 & 3.897e-01 & 8 \\
3.827e-02 & 4.021e-01 & 8 \\
4.388e-02 & 4.226e-01 & 8 \\
5.599e-02 & 4.619e-01 & 8 \\
7.054e-02 & 5.023e-01 & 8 \\
7.474e-02 & 5.123e-01 & 8 \\
8.388e-02 & 5.347e-01 & 8 \\
9.639e-02 & 5.623e-01 & 8 \\
1.467e-01 & 6.502e-01 & 8 \\
2.096e-01 & 7.309e-01 & 8 \\
2.627e-01 & 7.854e-01 & 8 \\
3.267e-01 & 8.368e-01 & 8 \\
3.927e-01 & 8.802e-01 & 8 \\
5.547e-01 & 9.589e-01 & 8 \\
6.773e-01 & 9.766e-01 & 8 \\
8.272e-01 & 9.618e-01 & 8 \\
9.960e-01 & 9.989e-01 & 8 \\\hline \hline
\end{longtable}

\end{document}